\makeatletter
\newcommand{\dontusepackage}[2][]{%
  \@namedef{ver@#2.sty}{9999/12/31}%
  \@namedef{opt@#2.sty}{#1}}
\makeatother
\dontusepackage{subfigure}
\documentclass[endfloat,manuscript]{geophysics}
\usepackage{amssymb,amsmath,amsbsy,bm}
\usepackage{fixltx2e} 
\IfFileExists{upquote.sty}{\usepackage{upquote}}{}
\usepackage[T1]{fontenc}
\usepackage[utf8]{inputenc}
\IfFileExists{microtype.sty}{\usepackage{microtype}}{}
\usepackage{natbib}
\usepackage{listings}
\usepackage{graphicx}
\makeatletter
\def\ScaleIfNeeded{%
  \ifdim\Gin@nat@width>\linewidth
    \linewidth
  \else
    \Gin@nat@width
  \fi
}
\makeatother
\let\Oldincludegraphics\includegraphics
{
 \catcode`\@=11\relax
 \gdef\includegraphics{\@ifnextchar[{\Oldincludegraphics}{\Oldincludegraphics[width=\ScaleIfNeeded]}}
}
\usepackage{caption}
\usepackage{float}
\usepackage{subfig}
\usepackage{algorithm} 
\let\scholmdAlgorithm\algorithm
\let\endscholmdAlgorithm\endalgorithm
\let\algorithm\relax \let\endalgorithm\relax

{
 \catcode`\*=11\relax
 \global\let\scholmdAlgorithm*\algorithm*
 \global\let\endscholmdAlgorithm*\endalgorithm*
 \global\let\algorithm*\relax 
 \global\let\endalgorithm*\relax
}

\righthead{SPLS-RTM with source estimation}

\usepackage[unicode=true]{hyperref}
\hypersetup{breaklinks=true,
            bookmarks=true,
            pdfauthor={},
            pdftitle={Time-domain sparsity promoting least-squares reverse time migration with source estimation},
            colorlinks=true,
            citecolor=black,
            urlcolor=blue,
            linkcolor=black,
            pdfborder={0 0 0}}
\newcommand{\mvec}{\mathbf{m}}
\newcommand{\uvec}{\mathbf{u}}
\newcommand{\dvec}{\mathbf{d}}
\newcommand{\xvec}{\mathbf{x}}

\newcommand{\qvec}{\mathbf{q}}
\newcommand{\dm}{\delta\mvec}
\newcommand{\dd}{\delta\dvec}

\newcommand{\nsrc}{n_{\text{s}}}

\newcommand{\Pmat}{\mathbf{P}}
\newcommand{\Jmat}{\mathbf{J}}

\newcommand{\Cmat}{\mathbf{C}}
\def\argmin{\mathop{\rm arg\,min}}

\newcommand{\minim}{\mathop{\mathrm{minimize}}}

\title{Time-domain sparsity promoting least-squares reverse time migration with
source estimation}
\author{Mengmeng Yang\textsuperscript{1*\#}, Zhilong Fang\textsuperscript{2*},
Philipp Witte\textsuperscript{3} and Felix J.
Herrmann\textsuperscript{1,3}\\\textsuperscript{1} School of Earth and
Atmospheric Sciences, Georgia Institute of
Technology\\\textsuperscript{2} Department of Mathematics, Massachusetts
Institute of Technology\\\textsuperscript{3} School of Computational
Science and Engineering, Georgia Institute of Technology\\ * Equally
contributed\\\# To whom correspondence should be addressed:
myang64@gatech.edu}
\date{}

\begin{document}
\maketitle

\thispagestyle{empty}

\begin{abstract}

Least-squares reverse time migration is well-known for its capability to
generate artifact-free true-amplitude subsurface images through fitting
observed data in the least-squares sense. However, when applied to
realistic imaging problems, this approach is faced with issues related
to overfitting and excessive computational costs induced by many
wave-equation solves. The fact that the source function is unknown
complicates this situation even further. Motivated by recent results in
stochastic optimization and transform-domain sparsity-promotion, we
demonstrate that the computational costs of inversion can be reduced
significantly while avoiding imaging artifacts and restoring amplitudes.
While powerfull, these new approaches do require accurate information on
the source-time function, which is often lacking. Without this
information, the imaging quality deteriorates rapidly. We address this
issue by presenting an approach where the source-time function is
estimated on the fly through a technique known as variable projection.
Aside from introducing negligible computational overhead, the proposed
method is shown to perform well on imaging problems with noisy data and
problems that involve complex settings such as salt. In either case, the
presented method produces high resolution high-amplitude fidelity images
including an estimates for the source-time function. In addition, due to
its use of stochastic optimization, we arrive at these images at roughly
one to two times the cost of conventional reverse time migration
involving all data.


\end{abstract}

\section{Introduction}\label{introduction}

Reverse-time migration (RTM) is a popular wave equation-based seismic
imaging methodology where the inverse of the linearized Born scattering
operator is approximated by applying its adjoint directly to the
observed reflection data
\citep{baysal1983reverse, whitmore1983iterative}. Because the adjoint
does not equal the pseudo inverse, conventional RTM produces images with
incorrect amplitudes. Among the factors that contribute to low-fidelity
amplitudes, the imprint of the temporal bandwidth limitation of the
typically unknown source wavelet features prominently and so does the
fact that the Born scattering operator is not inverted. To overcome
these issues, we formulate our imaging problem as a linear least-squares
inversion problem where the difference between observed and predicted
data is minimized in the $\ell_2$-norm
\citep{schuster1993least, nemeth1999least, dong2012least, zeng2014least}.
While least-squares migration is a powerful technique, its succesful
application to industry-scale problems is hampered by three key issues.
First, iterative demigrations (i.e.~Born modeling) and migrations become
computationally prohibitively expensive when carried out over all shots.
Second, we run the risk of overfitting the data when minimizing the
$\ell_2$-norm of the data residual. This overfitting may introduce
noise-related artifacts in inverted images. Third, while the source
location is generally well known, the temporal source function is often
not known accurately. Because imaging relies on knowing the source
function, this may have a detrimental effect on the image and makes it
necessary to come up with source estimation methodology. Since we carry
out our imaging iteratively, we propose to estimate the wavelet on the
fly as we build up the image.

We address the issue of computational feasibility by combining
techniques from stochastic optimization
\citep{van2011seismic, haber2012effective, powell2014clearing},
curvelet-domain sparsity-promotion \citep{herrmann2012efficient}, and
online convex optimization \citep{lorenz2014linearized} with linearized
Bregman. Stochastic optimization allows us to work with small random
subsets of shots, which limits the number of wave equation
solves---i.e., passes through the data. Convergence is guaranteed
\citetext{\citealp{herrmann2015fast}; \citealp[
]{yang2016time}; \citealp{witte2019compressive}} by replacing the
$\ell_1$-norm, by an elastic net consisting of a strongly convex
combination of $\ell_1$- and $\ell_2$-norm objectives. Inclusion of the
$\ell_2$-norm results in a greatly simplified algorithm involving
linearized Bregman iterations, which corresponds to gradient descent on
the dual variable supplemented by a simple soft thresholding operation
\citep{cai2009convergence, yin2010analysis} with a threshold that is
fixed. We refer to this method as sparsity-promoting least-squares
reverse-time migration (SPLS-RTM).

In addition to the high computational cost, the lack of accurate
knowledge on the unknown temporal source signature may also adversely
affect the performance of the inversion. Errors in the source signature
lead to erroneous residuals, which in turn result in inaccurately imaged
reflectors, which now may be positioned wrongly or may have the wrong
amplitude or phase. To mitigate these errors, we need an embedded
procedure where the source signature is updated along with the image
during the inversion
\citep{pratt1999seismic, aravkin2012source, aravkin2013sparse, fang2018source}
using a technique known as variable projection
\citep{rickett2013variable, van2014reply}. For time-harmonic imaging,
variable projection involves the estimation of the source function by
solving a least-squares problem for each frequency separately. Since the
unknown for each frequency in that case is a single complex-valued
variable, this process is simple and has resulted in an accurate
estimation and compensation for the source-time function
\citetext{\citealp[ ]{tu2013fast}; \citealp{fang2018source}}.
Unfortunately, the situation is more complicated during imaging in the
time domain, where we have to estimate the complete source signature
during each iteration. For this purpose, we integrate early work by
\citet{yang2016time} and inverse-scattering method
\citep{witte2019compressive} and achieve an approach that is suitable
for realistic imaging scenarios that may include salt.

Our work is outlined as follows. First, we introduce the basic equations
for time-domain reverse time migration and least-squares reverse time
migration. To overcome the computational cost associated with the
latter, we introduce a stochastic optimization method with sparsity
promotion. This method is designed to provide an image at a fraction of
the cost. Next, we extend this approach so it includes on-the-fly source
estimation. This allows us to remove the requirement of the source
function. We conclude by presenting a number of synthetic case studies
designed to demonstrate robustness with respect to noisy data and to
complex imaging scenarios that include salt.

\section{Methodology}\label{methodology}

Since our approach hinges on cost-effective least-squares imaging, we
first introduce our formulation of sparsity-promoting least-squares
migration with stochastic optimization, followed by our approach to
on-the-fly source estimation during the iterations.

\subsection{From RTM to LS-RTM}\label{from-rtm-to-ls-rtm}

Reverse time migration derives from a linearization (see e.g.
\citet{mulder2004comparison}) with respect to the background squared
slowness. For the $i^\mathrm{th}$ source this linearization reads
\begin{equation}
    \dd_{i} = \mathbf{F}_{i}(\mvec_0 + \dm, \qvec) - \mathbf{F}_{i}(\mvec_0, \qvec) \approx \Jmat_{i}(\mvec_0, \qvec) \dm,
\label{BornModeling}
\end{equation}
 where the vectors $\dm$, $\qvec$, and $\dd$ denote the model
perturbation, the source-time function, and the corresponding data
perturbation, respectively. We model the data for $n_t$ time samples
over a time interval of $T\,\mathrm{s}$. The number of receivers is
$n_r$ so a single shot record is of size $n_t\times n_r$. The nonlinear
forward modeling operator $\mathbf{F}_{i}(\mvec, \qvec)$ for the $i^\mathrm{th}$
source location involves the solution of the discretized acoustic wave
equation
\begin{equation}
\begin{aligned}
    \left (\mvec\odot\frac{\partial^2}{\partial t^{2}} - \Delta\right)\uvec_{i}&=\Pmat_{\text{s},i}^{\top}\qvec, \\
    \Pmat_{\text{r},i} \uvec_{i} &= \dvec_{i},
\end{aligned}
\label{AcousticWave}
\end{equation}
 parameterized by the squared slowness collected in the vector $\mvec$
(for simplicity, we kept the density constant and we used the symbol
$\odot$ to denote elementwise multiplication.) The symbol $\Delta$
represents the discretized Laplacian and the linear operator
$\Pmat_{\text{r},i}$ restricts the wavefield for the $i^\mathrm{th}$
source to the corresponding receiver locations, while the linear
operator $\Pmat_{\text{s},i}^{\top}$ injects the source time function at
the location of the $i^\mathrm{th}$ source in the computational grid.
The Jacobian $\Jmat_{i}(\mvec_0, \qvec)$ is known as the Born modeling
operator and is given by the derivative of $\mathbf{F}_i(\mvec, \qvec)$ at the
point of $\mvec_0$. Applying the Jacobian $\Jmat_{i}(\mvec_0, \qvec)$ to
the model perturbation $\dm$ requires the solution of the following
linearized equation:
\begin{equation}
\begin{aligned}
    \left(\mvec_0\odot\frac{\partial^2}{\partial t^{2}} - \Delta\right)\delta\uvec_{i}& = -\frac{\partial^2}{\partial t^2} \big( \delta\mvec\odot\uvec_{i} \big), \\
    \Pmat_{\text{r},i} \delta\uvec_{i} &= \dd_{i},
\end{aligned}
\label{BornWave}
\end{equation}
 where the vector $\delta\uvec_{i}$ corresponds to the wavefield
perturbation for the $i^\mathrm{th}$ source.

The goal of seismic imaging is to estimate the model perturbations from
observed data. We can expect this reconstruction process to be
successful in situations where the above linear approximation is
accurate---i.e., the background velocity model needs to be sufficiently
accurate, which we assume it is. We also need accurate knowledge on the
source function, an important aspect we will address below.

While the above linearization allows us to create an image via
\begin{equation}
\dm_{\text{RTM}}=\sum_{i=1}^{n_s} \Jmat^\top_{i}\dd_i
\label{migration}
\end{equation}
 with $n_s$ the number of shots, the adjoint of the Jacobian (denoted by
the symbol $^\top$) does not correspond to its inverse and
$\dm_{\text{RTM}}$ will suffer from wavelet side lobes and inaccurate
and unbalanced amplitudes
\citep{mulder2004comparison, bednar2006two, hou2016accelerating}. Unlike
RTM (Equation~\ref{migration}), LS-RTM
\citep{aoki2009fast, herrmann2012efficient, tu2015fast} reconstructs the
model perturbation by computing the pseudo-inverse of the Born modeling
operator, which can significantly mitigate these defects. LS-RTM
iteratively solves the following least squares data-fitting problem:
\begin{equation}
\minim_{\dm}\frac{1}{2}\sum_{i=1}^{\nsrc}\|\Jmat_{i}(\mvec_0,\qvec)\dm-\dd_{i}\|^2.
\label{Lineardata}
\end{equation}
 Compared to Equation~\ref{migration}, the above minimization requires
multiple evaluations of the Jacobian and its adjoint, which becomes
rapidly computationally prohibitive for large 2D and 3D imaging problems
as the number of sources $n_s$ grows. This in part explains the
relatively slow adaptation of least-squares reverse time migration
(cf.~Equation~\ref{Lineardata}) by industry. As we show below, we
overcome this problem by combining ideas from stochastic optimization
and sparsity promotion
\citep{herrmann2015fast, yang2016time, witte2019compressive}, which
allow us to obtain artifact-free images at the cost of two to three
passes through the data.

\subsection{Stochastic optimization with sparsity
promotion}\label{stochastic-optimization-with-sparsity-promotion}

As we mentioned above, the minimization of Equation~\ref{Lineardata}
over all $n_s$ shots is computationally prohibitively expensive. In
addition, the minimization is unconstrained and misses regularization to
battle the adverse effects of noise and the null space (missing
frequencies and finite aperture) associated with solving the
least-squares imaging problems of the type listed in
Equation~\ref{Lineardata}. To address these two problems, we combine
ideas from stochastic optimization, during which we only work on
randomized subsets of shots during each iteration, and ideas from
sparsity-promoting optimization designed to remove the imprint of the
null space and source subsampling related artifacts. As we have learned
from the field of compressive sensing
\citep{candes2006compressive, donoho2006compressed, candes2008introduction},
transform-domain sparsity promotion is a viable technique to remove
subsample related noise in imaging via
\begin{equation}
\begin{aligned}
  \minim_{\xvec} \quad &\|\xvec \|_{1}, \\
  \text{subject to } \,  &\sum_{i=1}^{n_s} \|\Jmat_{i}( \mvec_0, \mathbf{q}) \Cmat^{\top} \xvec - \delta \dvec_{i}\|_{2} \leq \sigma.
\end{aligned}
\label{BPDN}
\end{equation}
 In this formulation, known as the Basis Pursuit Denoise (BPDN,
\citet{chen2001atomic}) problem, we include the sparsity-promoting
$\ell_1$-norm as the objective on the curvelet coefficients $\xvec$ of
the image. These coefficients are related to the linearized data via the
adjoint of the curvelet transform $(\Cmat^{\top})$ and the above program
seeks to find the sparsest curvelet coefficient vector that matches the
data within the noise level $\sigma$. While the above problem is known
to produce high-fidelity results, its solution relies on iterations that
involve a loop over all $n_s$ shots.

Stochastic gradient descent \citep[e.g.][ in the context of seismic
inversion]{haber2012effective} is a widely used tool to make
unconstrained optimization problems of the type included in
Equation~\ref{Lineardata} computationally feasible by computing the
gradient of Equation~\ref{Lineardata} for randomized subsets of shots,
using a given a batch size that corresponds to the number of shots used
per iteration. This approach allows to minimize
Equation~\ref{Lineardata} in very few epochs, using only few passes
through data consisting of $n_s$ shot records, as long as the step
lengths adhere to certain conditions to guarantee convergence.
Unfortunately, this complicates the solution of BPDN. To avoid this
complication, we reformulate, following \citet{cai2009convergence},
Equation~\ref{BPDN} by replacing its convex $\ell_1$-norm objective by
the strongly convex objective involving
\begin{equation}
\begin{aligned}
\minim_{\mathbf{x}} \ & \lambda_{1} \Vert\mathbf{x} \Vert_1 + \frac{1}{2} \Vert\mathbf{x} \Vert^{2}_{2} \\
 \text{subject to} \ &\sum_{i=1}^{n_s} \Vert \Jmat_{i}( \mathbf{m}_0, \mathbf{q}) \mathbf{C}^{\top} \mathbf{x} - \delta \mathbf{d}_{i}\Vert_{2} \leq \sigma
\end{aligned}
\label{PlainLB}
\end{equation}
 with the estimate for the image given by
$\delta\mathbf{\hat{m}}=\mathbf{C}^{\top} \mathbf{\hat{x}}$ where
$\mathbf{\hat{x}}$ is the minimizer of the above optimization problem.
The mixed objective in this problem is known as an elastic net in
machine learning, which offers convergence guarantees (see
\citet{lorenz2014linearized}) in situations where we work during each
iteration with different randomized subsets of shots indexed by
$\mathcal{I}_k\subset [1\cdots n_s]$ with cardinality
$|\mathcal{I}|=n_s^\prime\ll n_s$. We choose these subsets without
replacement.

For $\lambda\rightarrow\infty$, which in practice means $\lambda$ large
enough, iterative solutions of Equation~\ref{PlainLB} as summarized in
Algorithm~\ref{alg:LSM1} converge to the solution of
Equation~\ref{BPDN}, even in situations where we work with randomized
subsets of shots. Compared to iterative solutions of
Equation~\ref{BPDN}, the iterations (lines 7--8 in
Algorithm~\ref{alg:LSM1}) correspond to iterative thresholding with a
fixed threshold $\lambda$ on the dual variable ($\mathbf{z_k}$) with a
dynamic step length given by
$t_k = \Vert \mathbf{A}_k \mathbf{x}_k - \mathbf{b}_k\Vert^{2}_{2} / \Vert \mathbf{A}_k^{\top} (\mathbf{A}_k \mathbf{x}_k - \mathbf{b}_k)\Vert^{2}_{2}$
\citep{lorenz2014linearized}. During each iteration, known as
linerarized Bregman iterations, the residual is projected onto an
$\ell_2$-norm ball with the radius $\sigma$ through a projection
operator $\mathcal{P}_\sigma$. To avoid too many iterations, we set the
threshold $\lambda$, related to the the tradeoff between the $\ell_1$
and $\ell_2$-norm objectives in Equation~\ref{PlainLB}, to a value that
is not too large---i.e., typically proportional to the maximum of
$|\mathbf{z}_k|$ at the first iteration ($k=1$). As reported by
\citet{yang2016time} and \citet{witte2019compressive}, high quality
images can be obtained running Algorithm~\ref{alg:LSM1} for a few epochs
as long as the source time function $\mathbf{q}$ and background velocity
model are sufficiently accurate. As we will show below, the background
velocity model also needs to be smooth so the tomography-related imaging
is avoided.

\begin{scholmdAlgorithm}
~~~1.~Initialize~$\mathbf{x}_0 = \mathbf{0}$,~$\mathbf{z}_0 = \mathbf{0}$,~$\mathbf{q}$,~$\lambda_{1}$,~batchsize~$n^\prime_{s} \ll n_s$\\\hspace*{0.333em}\hspace*{0.333em}\hspace*{0.333em}2.~\textbf{for}~~$k=0,1, \cdots$\\\hspace*{0.333em}\hspace*{0.333em}\hspace*{0.333em}3.~~~~~Randomly~choose~shot~subsets~$\mathcal{I}_k \subset [1 \cdots n_s],\, \vert \mathcal{I} \vert = n^\prime_{s}$\\\hspace*{0.333em}\hspace*{0.333em}\hspace*{0.333em}4.~~~~~$\mathbf{A}_k = \{\Jmat_i ( \mathbf{m}_0,\mathbf{q} ) \mathbf{C}^{\top}\}_{i\in\mathcal{I}_k}$\\\hspace*{0.333em}\hspace*{0.333em}\hspace*{0.333em}5.~~~~~$\mathbf{b}_k = \{\mathbf{\delta d}_i\}_{i\in\mathcal{I}_k}$\\\hspace*{0.333em}\hspace*{0.333em}\hspace*{0.333em}6.~~~~~$t_k = \Vert \mathbf{A}_k \mathbf{x}_k - \mathbf{b}_k\Vert^{2}_{2} / \Vert \mathbf{A}_k^{\top} (\mathbf{A}_k \mathbf{x}_k - \mathbf{b}_k)\Vert^{2}_{2}$\\\hspace*{0.333em}\hspace*{0.333em}\hspace*{0.333em}7.~~~~~$\mathbf{z}_{k+1} = \mathbf{z}_k - t_{k} \mathbf{A}^{\top}_k \mathcal{P}_\sigma (\mathbf{A}_k\mathbf{x}_k - \mathbf{b}_k)$\\\hspace*{0.333em}\hspace*{0.333em}\hspace*{0.333em}8.~~~~~$\mathbf{x}_{k+1}=S_{\lambda_{1}}(\mathbf{z}_{k+1})$\\\hspace*{0.333em}\hspace*{0.333em}\hspace*{0.333em}9.~\textbf{end}\\\hspace*{0.333em}\hspace*{0.333em}10.~\textbf{Output:}~$\hat{\dm}=\mathbf{C}^\top\mathbf{x}_{k+1}$\\\hspace*{0.333em}\hspace*{0.333em}\hspace*{0.333em}note:~$S_{\lambda_{1}}(\mathbf{z}_{k+1})=\mathrm{sign}(\mathbf{z}_{k+1})\max\{ 0, \vert \mathbf{z}_{k+1} \vert - \lambda_{1} \}$\\\hspace*{0.333em}\hspace*{0.333em}\hspace*{0.333em}\hspace*{0.333em}\hspace*{0.333em}\hspace*{0.333em}\hspace*{0.333em}\hspace*{0.333em}\hspace*{0.333em}\hspace*{0.333em}\hspace*{0.333em}\hspace*{0.333em}\hspace*{0.333em}$\mathcal{P}_{\sigma}(\mathbf{A}_k \mathbf{x}_k - \mathbf{b}_k) = \max\{ 0,1-\frac{\sigma}{\Vert \mathbf{A}_k \mathbf{x}_k -\mathbf{b}_k \Vert}\} \cdot (\mathbf{A}_k \mathbf{x}_k -\mathbf{b}_k)$
\caption{Linearized Bregman for SPLS-RTM}\label{alg:LSM1}
\end{scholmdAlgorithm}

\subsection{On-the-fly source
estimation}\label{on-the-fly-source-estimation}

In practice, we unfortunately do not have access to the source time
function $\mathbf{q}$ required by Algorithm~\ref{alg:LSM1}. Following
our earlier work on source estimation in time-harmonic imaging and
full-waveform inversion \citep{van2011seismic, tu2014fis}, we propose an
approach during which we estimate the source-time signature after each
model update by solving a least-squares problem that matches predicted
and observed data via a time-domain filter.

To keep our time-domain wave equation solvers with finite differences
numerically stable (In our implementation, we used Devito
(\url{https://www.devitoproject.org}) for our time-domain finite
difference simulations and gradient computations
\citep{devito-compiler, devito-api} and JUDI
(\url{https://github.com/slimgroup/JUDI.jl}) as an abstract linear
algebra interface to our Algorithms \citep{witte2018alf}), we introduce
an initial guess for the source time function $\mathbf{q}_0$ with a
bandwidth limited spectrum that is flat over the frequency range of
interest. Under some assumptions on the source time function, we can
write the true source time function as the convolution between the
initial guess and the unknown filter $\mathbf{w}$---i.e., we have
$\mathbf{q}=\mathbf{w}\ast\mathbf{q}_0$ where the symbol $\ast$ denotes
the temporal convolution. Because we assume one and the same source time
function for all shots, we can write
\begin{equation}
\Jmat_{i}(\mathbf{m}_0, \mathbf{w} \ast \mathbf{q}_{0}) = \mathbf{w} \ast  \Jmat_{i}(\mathbf{m}_0, \mathbf{q}_{0})
\label{SrcApp}
\end{equation}
 for all sources $i=1\cdots n_s$. In this expression, we make use of
linearity of the wave equation with respect to its source. To simplify
the notation, we also overload the temporal convolution (denoted by the
symbol $\ast$) to apply to all data---i.e.~all traces in the shot
records.

Based on the above relationship, we propose to solve for $\mathbf{w}$
after each linearized Bregman iteration (line 10 of
Algorithm~\ref{alg:LSM2}) via
\begin{equation}
\begin{aligned}
\min_{\mathbf{w}} \ &  \sum_{i\in\mathcal{I}_k} \Vert \mathbf{w} \ast \Jmat_{i}(\mathbf{m}_0, \mathbf{q}_{0}) \mathbf{C}^T \mathbf{x}  - \delta \mathbf{d}_i\Vert ^2_{2} + \Vert \mathbf{r}\odot (\mathbf{w} \ast \mathbf{q}_{0})  \Vert^2_{2}
\end{aligned}
\label{sub22}
\end{equation}
 with
$\mathcal{I}_k \subset [1 \cdots n_s], \vert \mathcal{I} \vert = n_{s}^\prime$
a randomly chosen shot subset of shot records.

To prevent overfitting while fitting the generated data
$\mathbf{\tilde{b}}_k$ at the $k\text{th}$ iteration to the observed
data $\mathbf{b}_k$, we include a penalty term $\mathbf{r}$ consisting
of an exponential weighting vector as given by:
\begin{equation}
  \mathbf{r}(t)= \nu + \log (1+e^{\alpha (t-t_0)}).
\label{Weight}
\end{equation}
 In this expression, the scalar $\alpha$ determines the rate of growth
after $t=t_0$. We choose $t_0$ such that oscillations related to
overfitting are suppressed after this time. This prevents overfitting
and ensures the filters $\mathbf{w}_k$ to be short such that the
estimated source time function $\mathbf{q}=\mathbf{w}_k\ast\mathbf{q}_0$
remains short as well. The weight parameter $\nu$ penalizes the energy
of the estimated source $\mathbf{q}$, which also helps to alleviate the
ill-conditioning of this sub-problem.

We summarize the different steps of our approach in
Algorithm~\ref{alg:LSM2} below. As earlier, we solve the
sparsity-promoting optimization problem via linearized Bregman
iterations, which now include a correlation (correlation denoted by the
symbol $\star$ is the adjoint of convolution) with the current estimate
for source time function correction ($\mathbf{w}_k$) in line 8. We
initialize the source time function correction with a discrete Delta
distribution ($\mathbf{w}_0=\mathbf{\delta}$). We refer to this method
with on-the-fly source estimation as sparsity-promoting LS-RTM with
source estimation (SPLS-RTM-SE).

\begin{scholmdAlgorithm}
~~~1.~Initialize~$\mathbf{x}_0 = \mathbf{0}$,~$\mathbf{w}_0=\mathbf{\delta}$,~$\mathbf{z}_0 = \mathbf{0}$,~$\mathbf{q}_0$,~$\mathbf{r}(t)$,~batchsize~$n_{s}^\prime\ll n_s$,~~$\mathbf{r}$\\\hspace*{0.333em}\hspace*{0.333em}\hspace*{0.333em}2.~\textbf{for}~~$k=0,1, \cdots$\\\hspace*{0.333em}\hspace*{0.333em}\hspace*{0.333em}3.~~~~~Randomly~choose~shot~subsets~~$\mathcal{I}_k \subset [1 \cdots n_s], \vert \mathcal{I} \vert = n_{s}^\prime$\\\hspace*{0.333em}\hspace*{0.333em}\hspace*{0.333em}4.~~~~~$\mathbf{A}_k = \{\Jmat_i ( \mathbf{m}_0,\mathbf{q}_{0} ) \mathbf{C}^{\top}\}_{i\in\mathcal{I}_k}$\\\hspace*{0.333em}\hspace*{0.333em}\hspace*{0.333em}5.~~~~~$\mathbf{b}_k = \{\mathbf{\delta d}_i\}_{i\in\mathcal{I}_k}$~~\\\hspace*{0.333em}\hspace*{0.333em}\hspace*{0.333em}6.~~~~~$\mathbf{\tilde{b}}_k = \mathbf{A}_k \mathbf{x}_k$\\\hspace*{0.333em}\hspace*{0.333em}\hspace*{0.333em}7.~~~~~$t_k = \Vert \mathbf{\tilde{b}}_k - \mathbf{b}_k\Vert^{2}_{2} / \Vert \mathbf{A}_k^{\top} (\mathbf{\tilde{b}}_k - \mathbf{b}_k)\Vert^{2}_{2}$\\\hspace*{0.333em}\hspace*{0.333em}\hspace*{0.333em}8.~~~~~$\mathbf{z}_{k+1} = \mathbf{z}_k - t_{k} \mathbf{A}^\top_{k} \Big( \mathbf{w}_k {\star} \mathcal{P}_\sigma (\mathbf{w}_k \ast \mathbf{\tilde{b}}_k - \mathbf{b}_k) \Big)$\\\hspace*{0.333em}\hspace*{0.333em}\hspace*{0.333em}9.~~~~~$\mathbf{x}_{k+1}=S_{\lambda}(\mathbf{z}_{k+1})$\\\hspace*{0.333em}\hspace*{0.333em}\hspace*{0.333em}10.~~~~$\mathbf{w}_{k+1} = \argmin_{\mathbf{w}} \Vert \mathbf{w} \ast \tilde{\mathbf{b}}_{k} - \mathbf{b}_{k}\Vert ^2_{2} + \Vert \mathbf{r}\odot (\mathbf{w} \ast \mathbf{q}_0) \Vert^2_{2}$\\\hspace*{0.333em}\hspace*{0.333em}11.~\textbf{end}\\\hspace*{0.333em}\hspace*{0.333em}12.~\textbf{Output:}~$\hat{\mathbf{q}}=\mathbf{w}_{k+1}\ast \mathbf{q}_0$~and~$\hat{\dm}=\mathbf{C}^\top\mathbf{x}_{k+1}$
\caption{LB for LS-RTM with source estimation}\label{alg:LSM2}
\end{scholmdAlgorithm}

\section{Numerical experiments}\label{numerical-experiments}

In this experiment section, we demonstrate the viability of our approach
by means of carefully designed synthetic examples. We start by showing
that linearized Bregman iterations with on-the-fly source estimation are
indeed able to jointly estimate the source and the sparse vector of
image curvelet coefficients. Next, we consider an imaging experiment on
the Marmousi model emphasizing the importance of including the source
function and the influence of noise. We conclude by introducing a
practical workflow that is capable of handling salt-related imaging
problems.

\subsection{Stylized example}\label{stylized-example}

While imaging with linearized Bregman (LB) iterations has resulted in
high-fidelity true-amplitude images in complex models
\citep{witte2019compressive}, the viability of this alternative
sparsity-promoting approach has not yet been verified in combination
with on-the-fly source estimation. For this purpose, we examine the
performance of LB with source estimation on a simplified stylized
example. As we can see, Equation~\ref{SrcApp} implies a bilinear
dependence of the reflected data on both the unknown filter $\mathbf{w}$
and the curvelet coefficient vector $\mathbf{x}$. It is well known that
this sort of bilinear dependence can give rise to ambiguities even
though the vector $\mathbf{x}$ is sparse.

We exemplify this seismic bilinear relationship be defining
$\mathbf{W}\mathbf{A}\mathbf{x}=\mathbf{b}$, where the matrix
$\mathbf{A} \in \mathcal{R}^{20000 \times 10000}$ is ill-conditioned,
with $\text{rank}(\mathbf{A})=500$. The sparse vector
$\mathbf{x}\in \mathcal{R}^{10000\times 1}$ has only $20$ random
non-zero elements. A block of the tall matrix,
$\mathbf{A}_i \in \mathcal{R}^{500\times 10000}, i\in [1\dots 40]$
serves as a proxy for the LB modeling operator $\mathbf{J}_i$ for the
$i^\mathrm{th}$ shot with only one single trace. We implement the
trace-by-trace convolution via a Toeplitz matrix defined in terms of the
filter $\mathbf{w}\in \mathcal{R}^{500\times 1}$ acting on each
$\mathbf{A}_i\mathbf{x}$. The multiplication of the convolution matrix
$\mathbf{W}\in \mathcal{R}^{20000 \times 20000}$ with $\mathbf{Ax}$
compactly represents the repeated convolutions of the filter with all
traces.

This example, designed to jointly invert for $\mathbf{x}$ and
$\mathbf{w}$, aims to exhibit the capability of our
Algorithm~\ref{alg:LSM2} to carry out seismic imaging and on-the-fly
source estimation. To demonstrate the effect of the penalty term in line
10 of Algorithm~\ref{alg:LSM2}, we compare sparsity-promoting solutions
for the fixed true wavelet to solutions with on-the-fly source
estimation with and without the additional penalty. During each
iteration, we randomly choose $10\%$ of the blocks of the tall matrix
$\mathbf{A}$ and we run five passes through the data in total. After
some parameter testing, we choose the following values for the penalty
parameters: $\lambda=1,\nu=1$ and $\alpha=8$. We find that different
choices for these penalty parameters have little effect on our inversion
results. Finally, the time parameter $t_0$ is set according to the
approximate duration of the filter $\mathbf{w}$, which in this case
corresponds to a Ricker wavelet since we choose $\mathbf{q}_0$ to be a
delta Dirac. We also initialize the filter $\mathbf{w}$ with a
normalized Dirac. Because of the well-known amplitude ambiguity inherent
to blind deconvolution problems, we normalize the $\ell_2$-norms of the
estimated source and reflectivity.

\begin{figure}
\centering
\subfloat[\label{fig_toy2_a}]{\includegraphics[width=0.500\hsize]{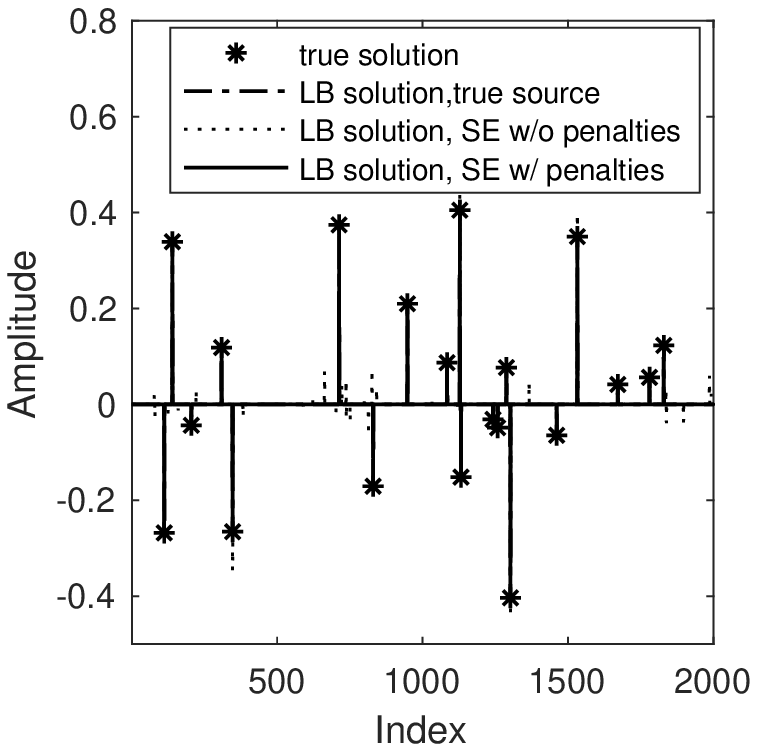}}
\subfloat[\label{fig_toy2_b}]{\includegraphics[width=0.500\hsize]{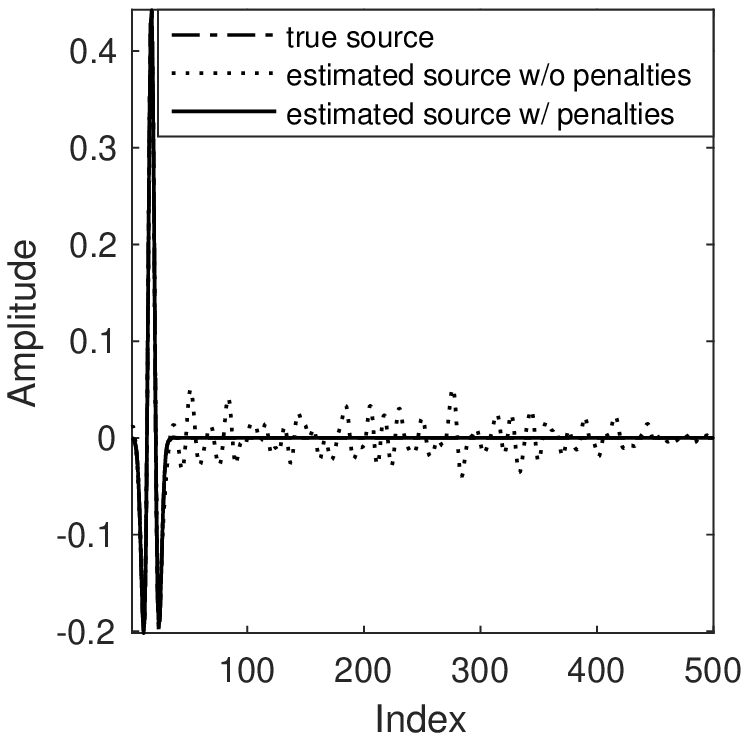}}
\\
\subfloat[\label{fig_toy2_c}]{\includegraphics[width=0.500\hsize]{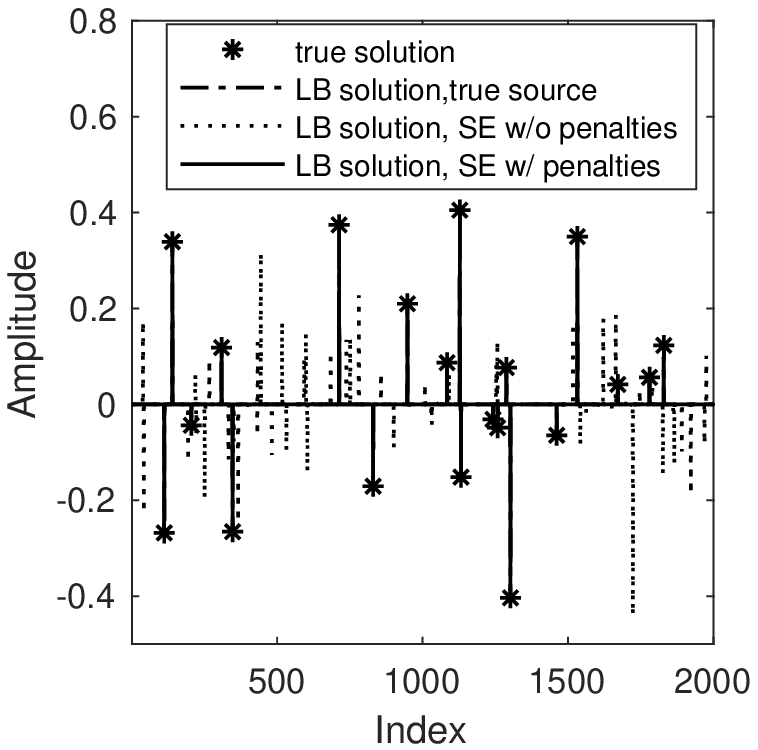}}
\subfloat[\label{fig_toy2_d}]{\includegraphics[width=0.500\hsize]{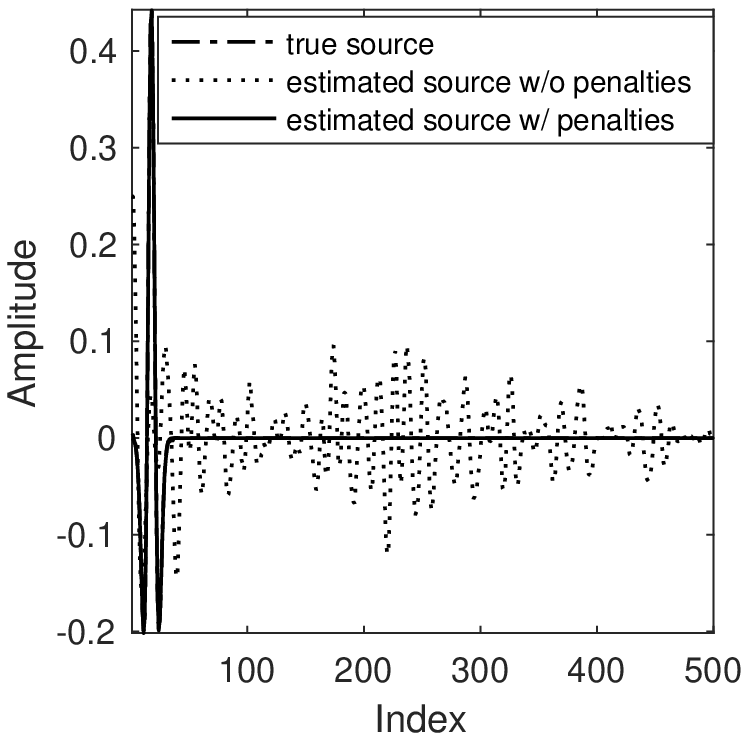}}
\caption{Comparison of solutions obtained with the LB iterations (see
Algorithm~\ref{alg:LSM2}) for a fixed true source (denoted by the
dash-dot line) and for on-the-fly source estimations with and without
penalties. We obtained results with five passes through the data. Our
method is capable of estimating the ``reflectivity'' (a) and ``source
function'' (b) after normalizing the $\ell_2$-norm. The proposed method
is also robust with respect to additive noise as we can see in (c) and
(d). We add $10\%$ Gaussian noise.}\label{fig_toy2}
\end{figure}

Pairs of estimated sparse ``reflectivities'' ($\hat{\dm}$) and source
functions ($\hat{\mathbf{q}}=\mathbf{w}_{final}\ast\mathbf{q}_0$) after
normalization are included in Figure~\ref{fig_toy2} . We can draw the
following conclusions from these results. First, for the noise-free
data, the LB iterations are able to recover the sparse ``reflectivity''
and source function well up to a constant single amplitude factor, which
we correct by normalizing its $\ell_2$-norm. Second, the estimated
source function and reflectivity become noisy (cf.~the dotted line in
Figure~\ref{fig_toy2}a and the dot line in Figure~\ref{fig_toy2}b ) when
we do not include a penalty enforcing the estimated filter to be short
in time. Finally, the method is robust with respect to noise as we can
see from Figures~\ref{fig_toy2}c and~\ref{fig_toy2}d where $10\%$
Gaussian noise is added. This result stresses the importance of
including the penalty.

\subsection{Experiments on the modified Marmousi
model}\label{experiments-on-the-modified-marmousi-model}

\begin{figure}
\centering
\subfloat[\label{fig_source_Marmousi_a}]{\includegraphics[width=0.500\hsize]{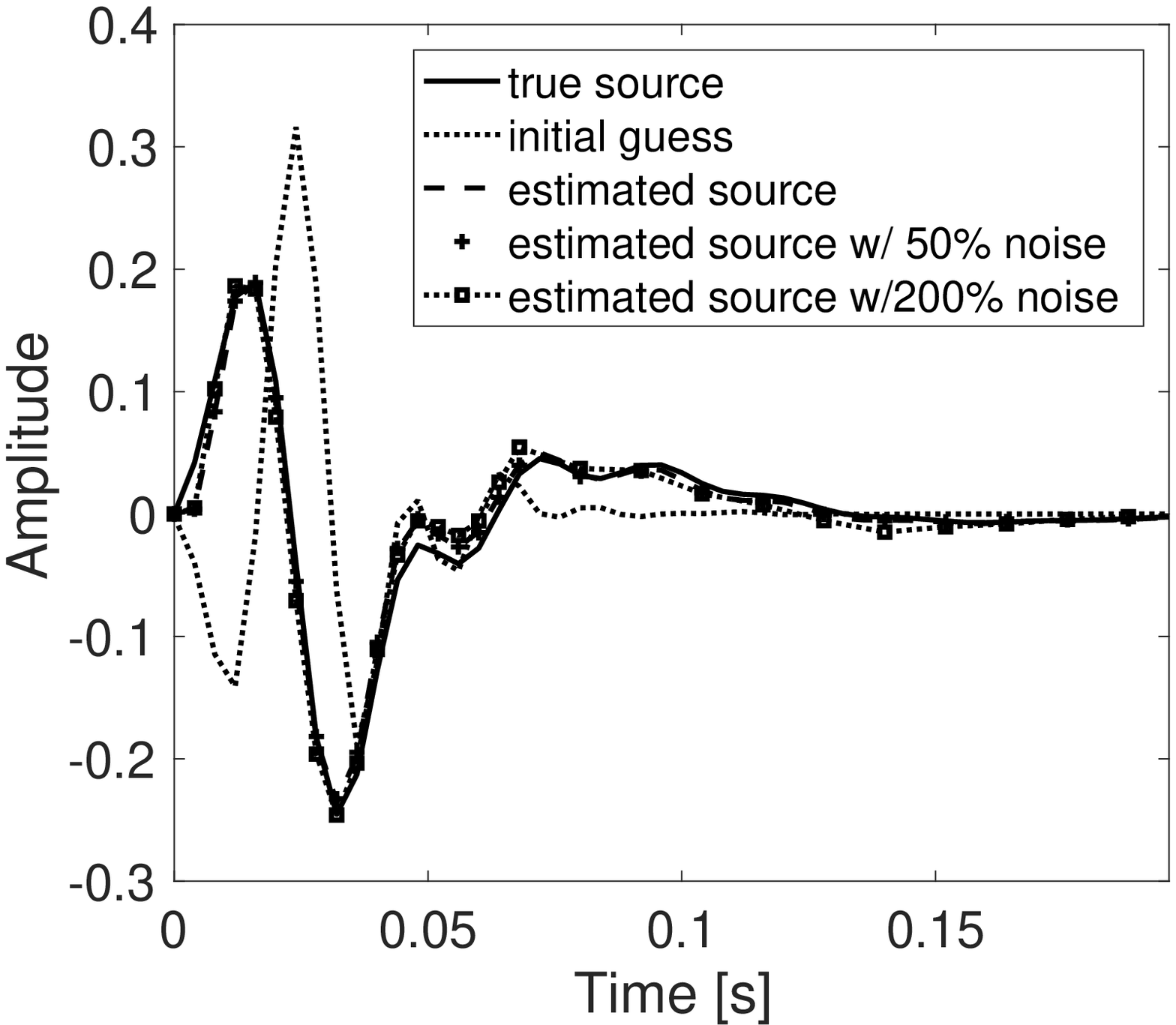}}  \hspace*{.4mm}
\subfloat[\label{fig_source_Marmousi_b}]{\includegraphics[width=0.500\hsize]{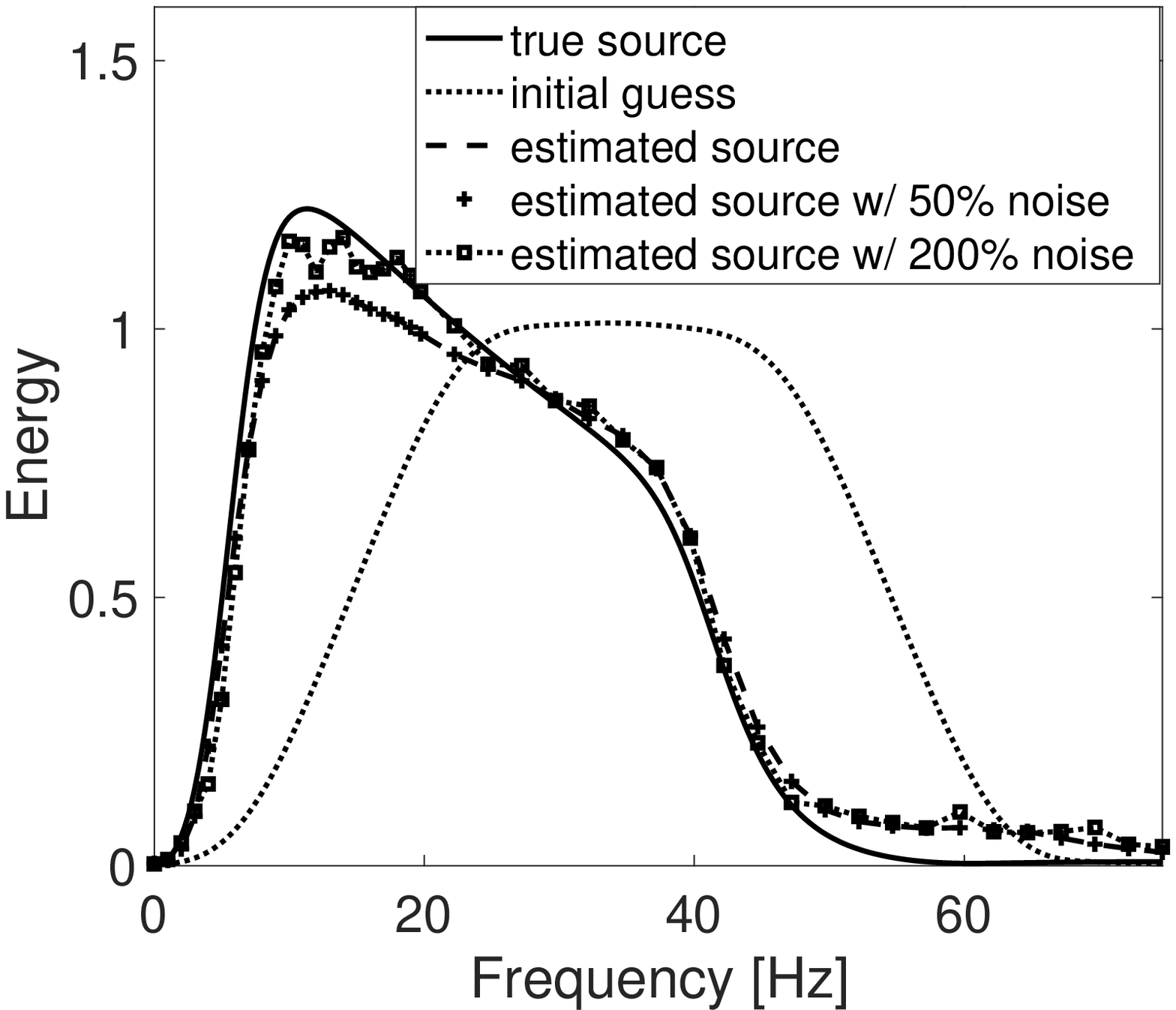}}
\caption{Comparison between true, initial, and estimated source time
functions ($\mathbf{q}_0$,
$\hat{\mathbf{q}}=\mathbf{w}_{final}\ast \mathbf{q}_0$) and their
associated amplitude spectra. (a) The time signatures and (b) the
frequency spectra. The estimated source time functions and spectra are
obtained from noise-free data and from data to which zero-centered
Gaussian noise is added with energy ranging from $50\%$ to $200\%$ of
the simulated linear data.}\label{fig_source_Marmousi}
\end{figure}

To illustrate the performance and robustness with respect to noise of
the proposed SPLS-RTM-SE method, we consider a model with complex
layered stratigraphy. We derive this imaging example from the well-known
synthetic Marmousi model \citep{brougois1990marmousi}, which is $3.2$ km
deep and $8.0$ km wide, with a grid size of $5\times 5$ m. To avoid
imaging artifacts, we use a background velocity that is kinematically
correct. We simulate $320$ equally spaced sources positioned at a depth
of $25$m. We use a minimum phase source time function with its
significant spectrum ranging from $10$ to $40$ Hz as shown in
Figure~\ref{fig_source_Marmousi}. We use this type of source to generate
linear data by applying the demigration operator
($\Jmat_i ( \mathbf{m}_0,\mathbf{q} ),\, i=1\cdots n_s$) to a bandwidth
limited medium perturbation $\dm$ given by the difference between two
smoothings of the true medium \citep{huang2016flexibly}. We record data
at $320$ equally spaced co-located receivers. To assess the sensitivity
to noise, we create two additional data sets by adding zero-centered
Gaussian noise whose energies are $50\%$ and $200\%$ of the simulated
linear data respectively.

Contrary to source estimation in the frequency domain, we need an
initial source function $\mathbf{q}_0$ for the source time function (see
Figure~\ref{fig_source_Marmousi_a} and~\ref{fig_source_Marmousi_b} where
the initial source time function and its amplitude spectrum are depicted
by dashed black lines). We need this initial source function to make
sure that the finite-difference propagators remain stable. To make sure
we do not exceed the valid frequency range of our simulations, we choose
the frequency band of the initial source time function broader than the
true one. To circumvent bias, we initialize the time function with a
flat amplitude spectrum between $20-50$ Hz. To allow for a realistic
scenario, we apply a phase shift to this initial guess making it mixed
phase and non symmetric .

To carry out the alternating inversion for the reflectivity and unknown
filter $\mathbf{w}$, we run Algorithm~\ref{alg:LSM2} for $40$ iterations
with a batch size of $8$---i.e., we use $8$ randomly selected sources
per iteration without replacement. The total number of wave equation
solves is equivalent to touching each shot only once---i.e., we make one
pass through the data. To improve the convergence of the inversion, we
employ preconditioners in both the data and model domains \citep[see][
for detail]{herrmann2009curvelet}. To remove the imprint of the
sources/receivers on the image, we also include a top mute to our
operators. Also, we apply a mute to the data to suppress the dominating
water bottom reflection and long offsets. Finally, we choose the
thresholding parameter $\lambda$ to be $10\%$ of the maximum value of
the first gradient to avoid unnecessary extra iterations resulting from
a threshold value that is too large or small.

\begin{figure}
\centering
\subfloat[\label{fig_LSRTM_Marmousi_a}]{\includegraphics[width=0.500\hsize]{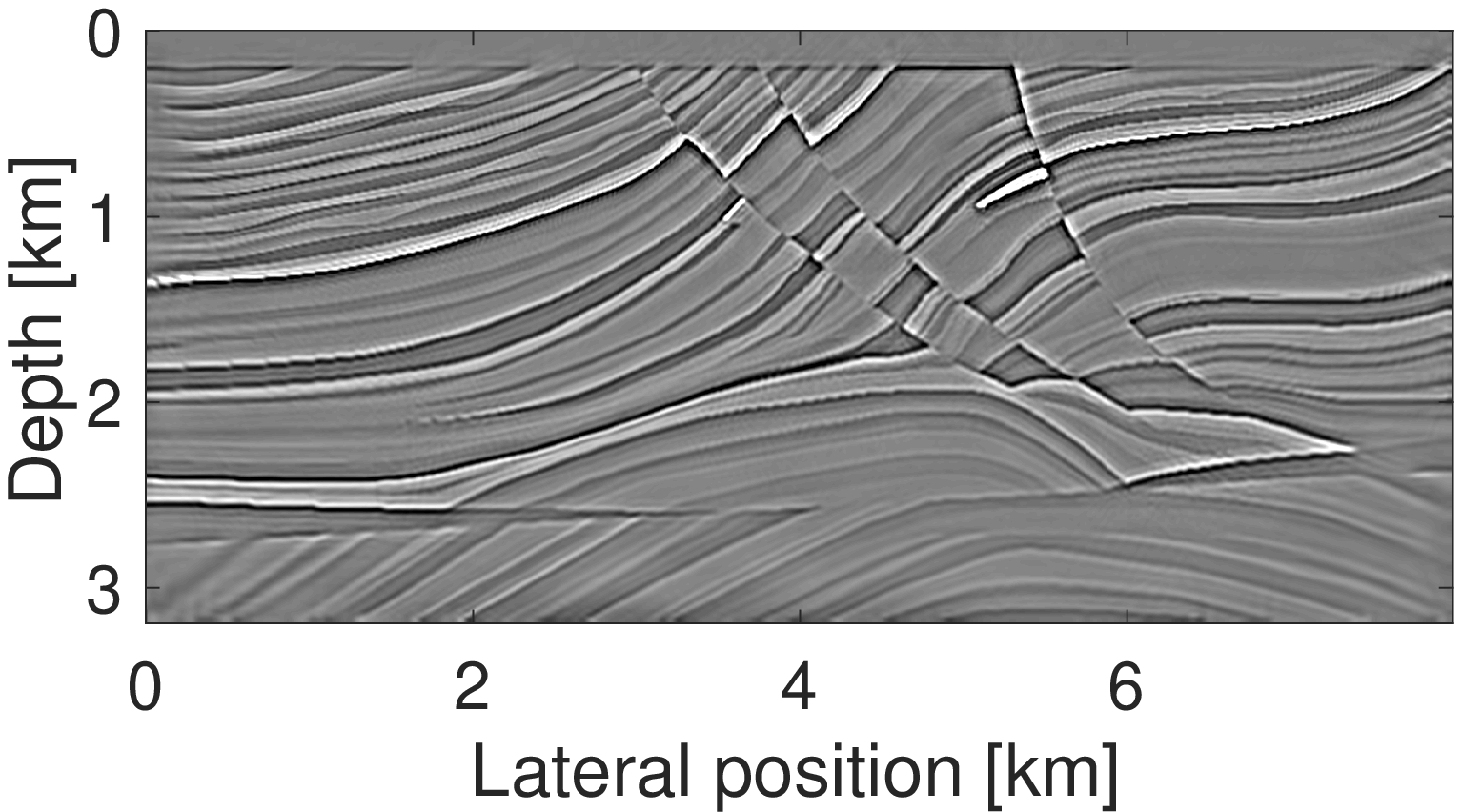}}\hspace*{.4mm}
\subfloat[\label{fig_LSRTM_Marmousi_b}]{\includegraphics[width=0.500\hsize]{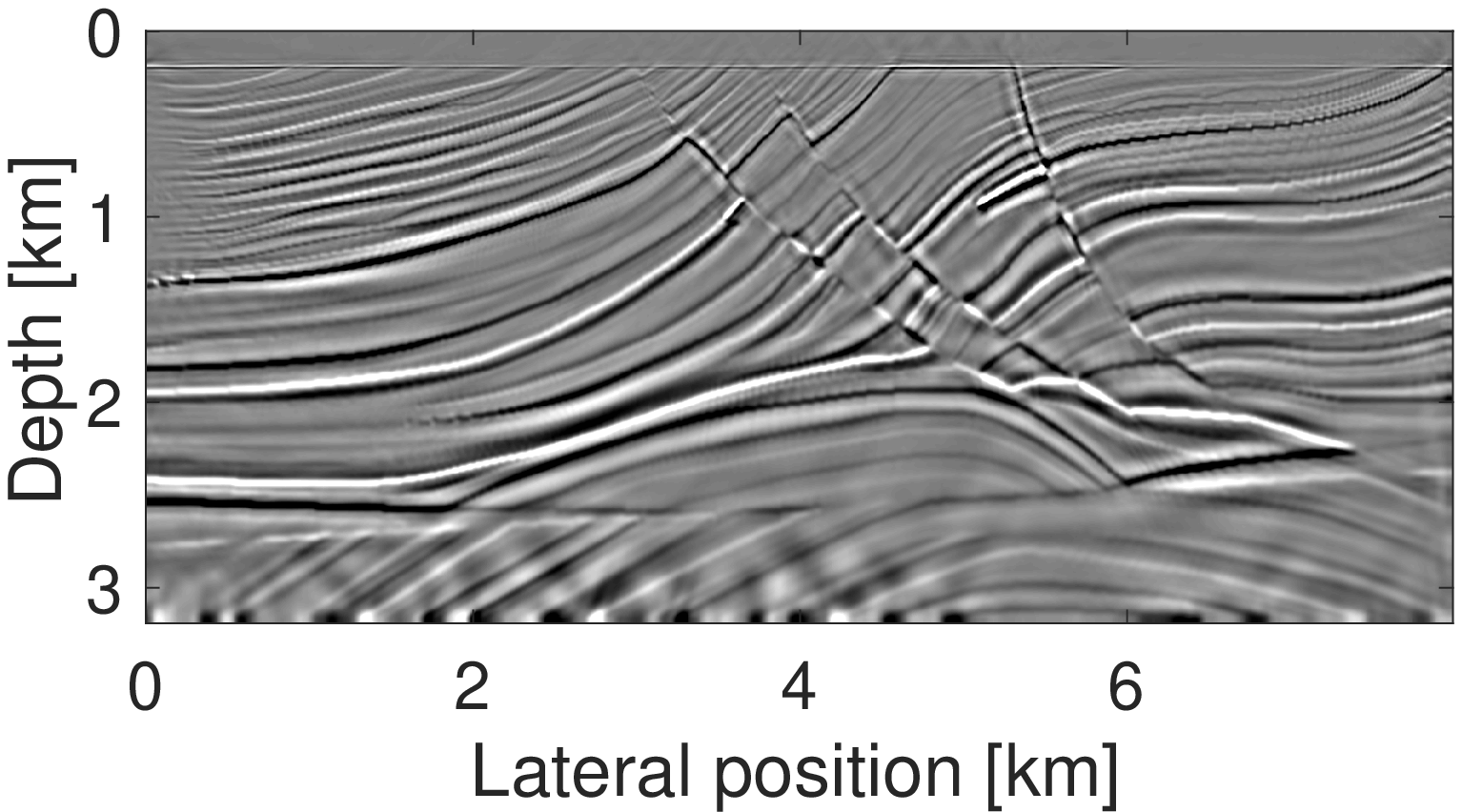}}
\caption{Inverted images with (a) the true source and (b) initial
source, generated by the Marmousi model.}\label{fig_LSRTM_Marmousi}
\end{figure}

\begin{figure}
\centering
\subfloat[\label{fig_LSRTM_Marmousi_estQ_a}]{\includegraphics[width=0.500\hsize]{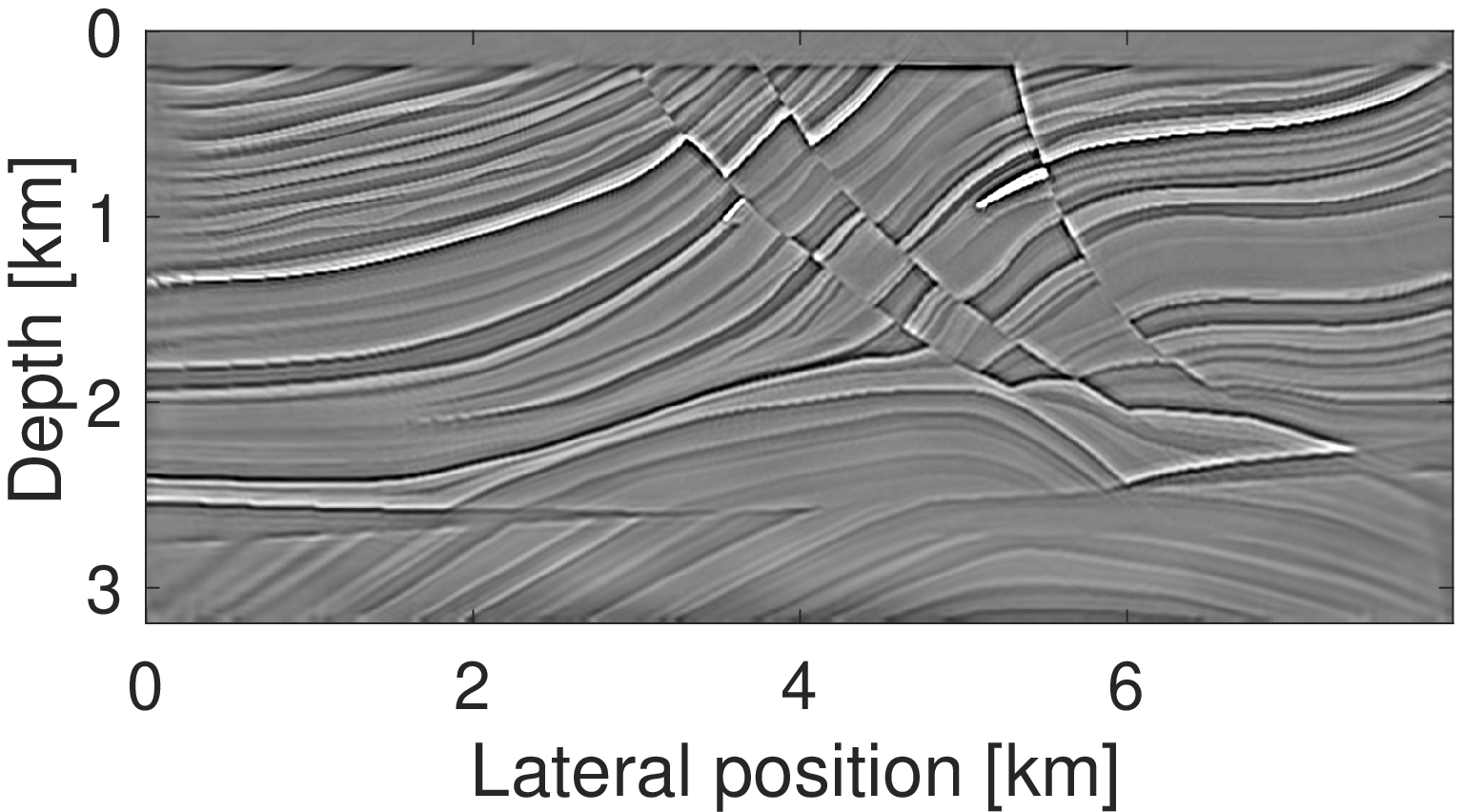}}\hspace*{.4mm}
\subfloat[\label{fig_LSRTM_Marmousi_estQ_b}]{\includegraphics[width=0.500\hsize]{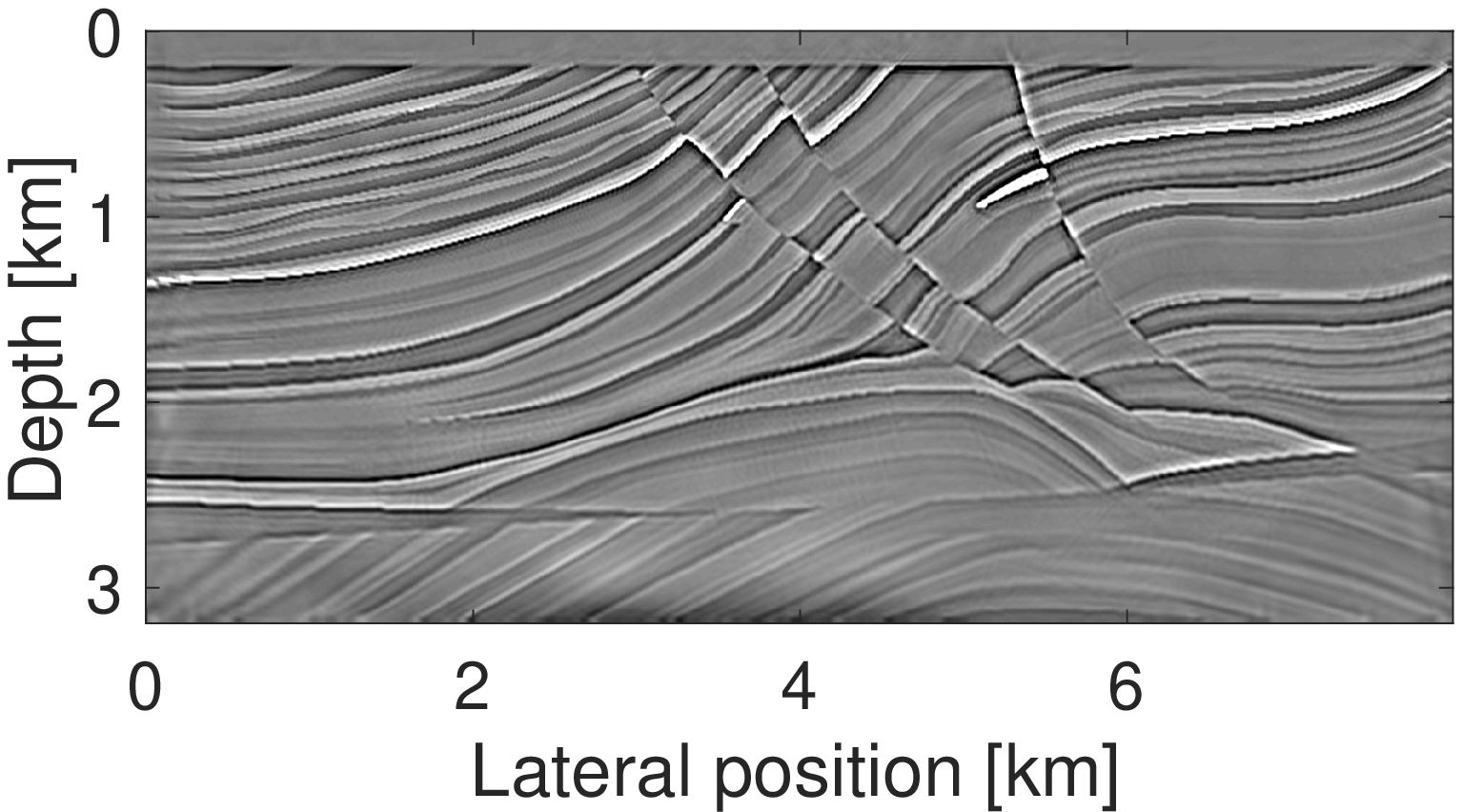}}
\\
\subfloat[\label{fig_LSRTM_Marmousi_estQ_c}]{\includegraphics[width=0.500\hsize]{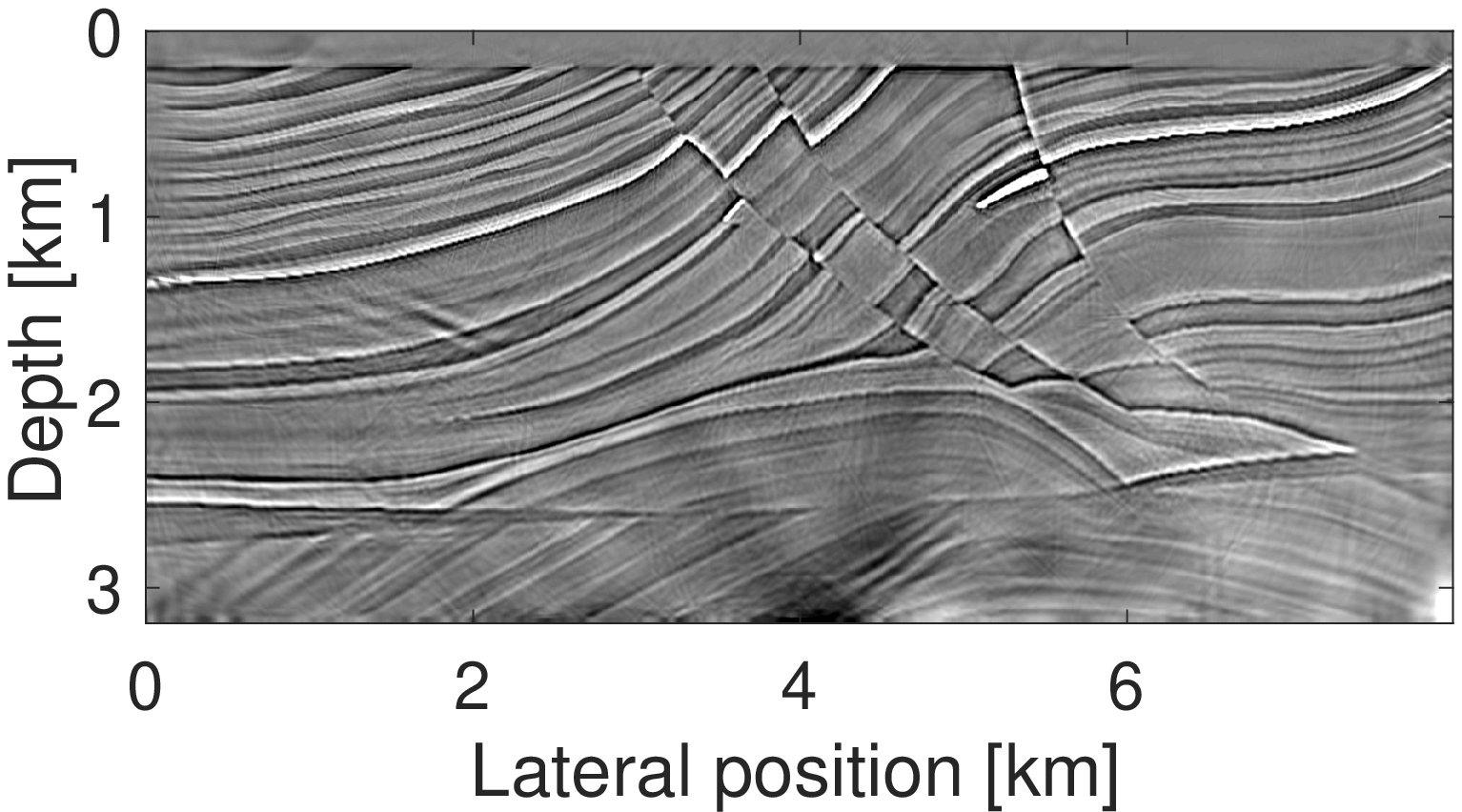}}
\caption{Inverted images with source estimation
(Algorithm~\ref{alg:LSM2}) for (a) noise-free data, (b) data with 50\%
noise and (c) data with 200\% noise, generated from the Marmousi
model.}\label{fig_LSRTM_Marmousi_estQ}
\end{figure}

The estimated source functions
$\hat{\mathbf{q}}=\mathbf{w}_{final}\ast\mathbf{q}_0$ and their
amplitude spectra after applying an $\ell_2$-norm normalization are
included in Figure~\ref{fig_source_Marmousi}. Overall we can see that
the source functions are well recovered despite the presence of noise.
For a small amount of noise, the estimated spectrum is the same as the
one obtained from the noise-free data while the source function obtained
from data with a high noise level is less smooth, but closer to the true
source function. Other than the fact that we are dealing with nonlinear
blind deconvolution, we do not have an explanation for this behavior.
While the noise dependence of the estimated source functions behaves
somewhat aberrant, the recovered reflectivity behaves as expected
(cf.~Figures~\ref{fig_LSRTM_Marmousi_a} and~\ref{fig_LSRTM_Marmousi_b}
for images obtained with the true source and with the initial guess and
images~\ref{fig_LSRTM_Marmousi_estQ_a}
--~\ref{fig_LSRTM_Marmousi_estQ_c} obtained with on-the-fly source
estimation for noise-free and noisy data.) We can make the following
observations from these experiments: first, it is important to image
with the correct source even when the data is noise free. While our
sparsity-promoting scheme is able to recover a high-resolution image
(see Figure~\ref{fig_LSRTM_Marmousi_a}) when the source function
corresponds to the true source, the image quality deteriorates rapidly
if the amplitude and phase spectra of the wavelet are wrong (see
Figure~\ref{fig_LSRTM_Marmousi_b}). Energy is no longer focussed and the
shape and locations of the imaged reflectors are off. However, the
results included in Figure~\ref{fig_LSRTM_Marmousi_estQ} demonstrate
that good results can be obtained when estimating the source function on
the fly. The estimated reflectivity depicted in
Figure~\ref{fig_LSRTM_Marmousi_estQ_a} is close to the reflectivity
obtained when we image with the true source function
(cf.~Figures~\ref{fig_LSRTM_Marmousi_a}
and~\ref{fig_LSRTM_Marmousi_estQ}). Moreover, the estimated images are,
as expected, relatively insensitive to noise in the data albeit the
imaged reflectivity for the high noise case somewhat deteriorated
(cf.~Figures~\ref{fig_LSRTM_Marmousi_estQ_a}
--~\ref{fig_LSRTM_Marmousi_estQ_c}). Contrary to the imaging result for
the wrong initial source function, the reflectors are positioned
correctly and have the correct phase, shape, and amplitude, even in
situations of substantial noise although at the expense of some
remaining low- and high-frequency artifacts. The latter are related to
the use of the curvelet transform and are to be expected. Overall, these
results confirm the robustness of our imaging algorithm in the situation
where there is significant noise in the data.

To arrive at the estimated images in
Figure~\ref{fig_LSRTM_Marmousi_estQ}, we set the penalty parameters
$\nu=1$ and $\alpha=8$ in Algorithm~\ref{alg:LSM2}. After the first
source estimation in the second iteration, we reset the coefficients
$\mathbf{z}$ and $\mathbf{x}$ to zero to avoid spending too many
iterations on correcting wrongly located reflectors from the first
iteration in which the initial guess of the source wavelet is used. In
addition to the visual quality of the estimated images, convergence
plots for the relative error for the data residual (the relative
$\ell_2$-norm error between the observed data and the demigrated data
for estimated reflectivity $\hat{\dm}$ convolved with the estimated
filter,$\frac{\|\mathbf{w}_k\ast\tilde{\mathbf{b}}_k-\mathbf{b}_k \|_2}{\|\mathbf{b}_k\|_2}$)
and the relative model error (the $\ell_2$-norm error between the true
reflectivity and the recovered
reflectivity,$\frac{\|\hat{\delta\mathbf{m}}_k -\delta\mathbf{m}\|_2}{\|\delta\mathbf{m}\|_2}$)
confirm our observation that Algorithm~\ref{alg:LSM2} is capable of
providing high quality images in the absence of precise knowledge on the
source function and in the presence of substantial noise. Our approach
arrives at these least-squares images at the cost of a single data pass.
Understandably, the algorithm starts off with a large relative residual
and model error due to the wrong initial guess for the source function.
As Algorithm~\ref{alg:LSM2} progresses, these relative errors continue
to decay and are comparable to the convergence plots for the true source
function. Because on-the-fly source estimation improves our ability to
adapt to the data, the relative data residual for the noise-free case
(dashed line) is even better then the relative error in case the source
function is known (solid line). While encouraging, these results are
obtained for a relatively simple imaging experiment and for data that is
obtained with linearized modeling via demigration. In other words, we
commit an inversion crime. In the next section, we will show that the
proposed method also performs well in more complicated settings with
nonlinear data.

\begin{figure}
\centering
\subfloat[\label{fig_residual_Marmousi_a}]{\includegraphics[width=0.500\hsize]{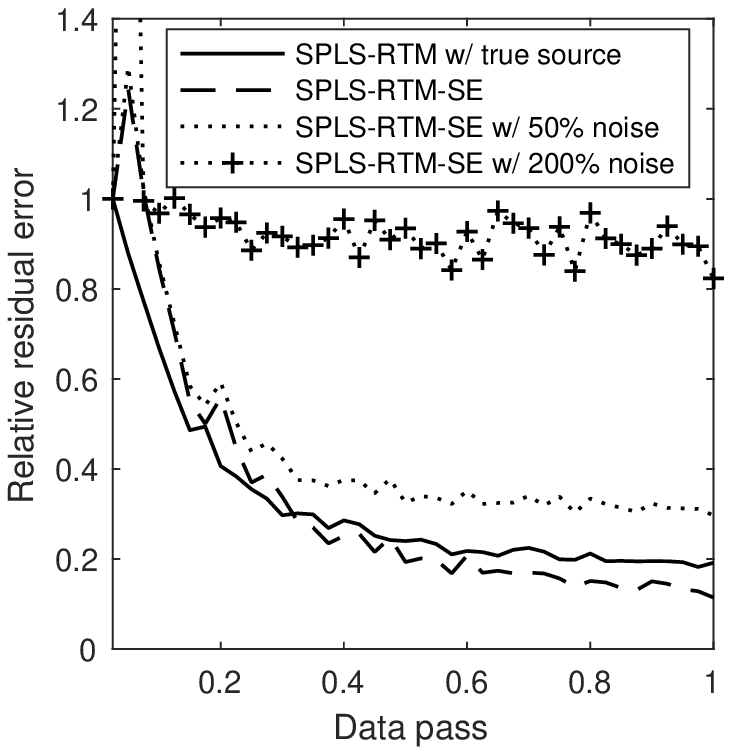}}\hspace*{.4mm}
\subfloat[\label{fig_residual_Marmousi_b}]{\includegraphics[width=0.500\hsize]{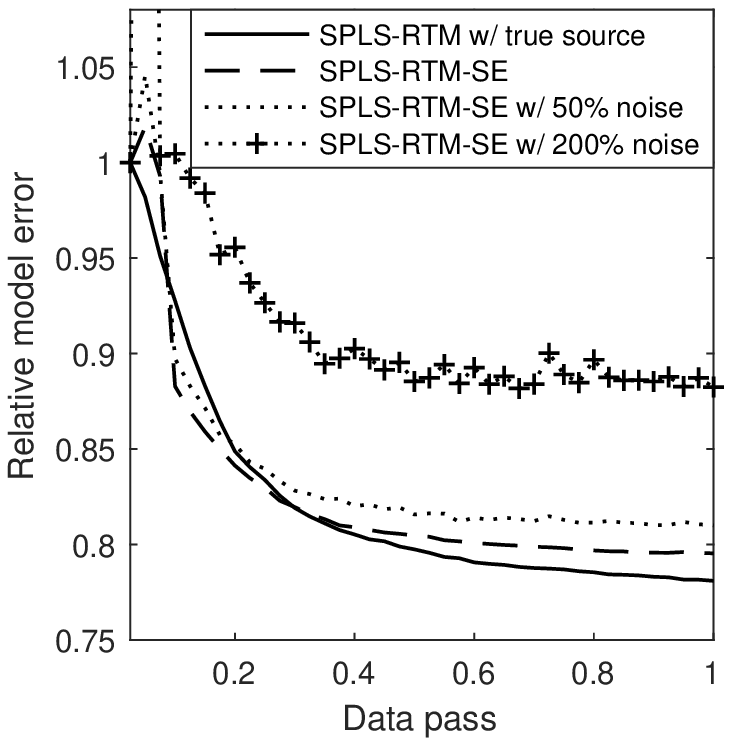}}
\caption{Convergence plots for the relative residual error (a)---i.e.,
the relative $\ell_2$-norm error between the observed data and the
demigrated data for estimated reflectivity $\hat{\dm}_k$ convolved with
the estimated filter $\mathbf{w}_k$, and the relative model error
(b)---i.e., the relative $\ell_2$-norm error between the true
reflectivity and the recovered
reflectivity.}\label{fig_residual_Marmousi}
\end{figure}

\subsection{Experiments on the Sigsbee
model}\label{experiments-on-the-sigsbee-model}

\begin{figure}
\centering
\subfloat[\label{fig_source_a}]{\includegraphics[width=0.500\hsize]{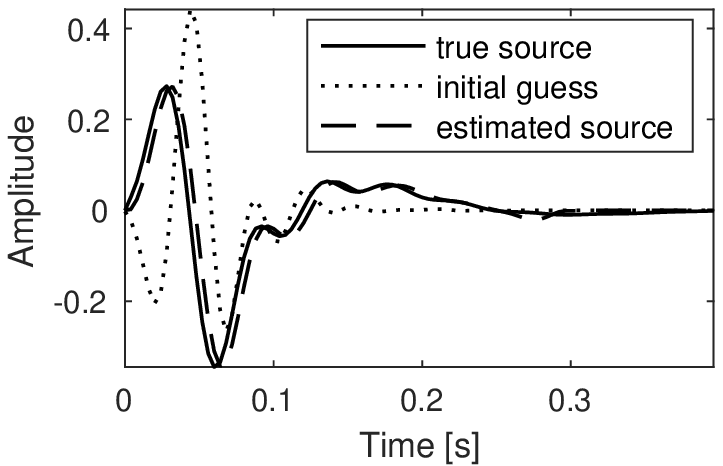}}\hspace*{.4mm}
\subfloat[\label{fig_source_b}]{\includegraphics[width=0.500\hsize]{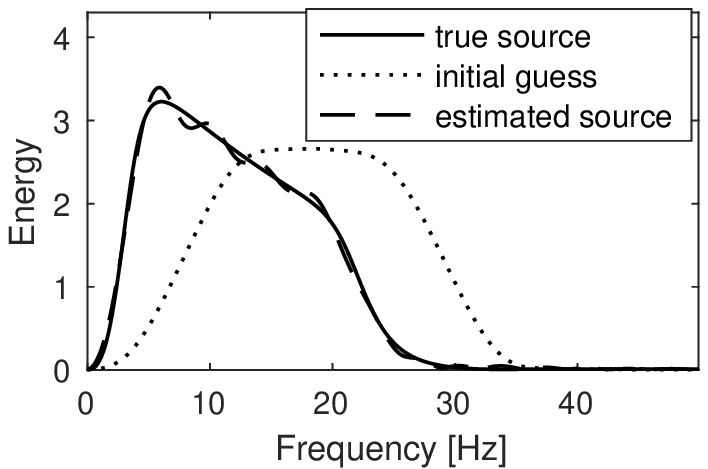}}
\caption{Comparison of true, initial, and estimated source time
functions ($\mathbf{q}_0$,
$\hat{\mathbf{q}}=\mathbf{w}_{final}\ast\mathbf{q}_0$) and their
associated amplitude spectra. (a) The time signatures and (b) the
frequency spectra. The estimated source time functions and spectra are
obtained during the first five iterations with the conventional imaging
condition.}\label{fig_source}
\end{figure}

Sparsity-promoting imaging algorithms such as SPLS-RTM
(Algoritm~\ref{alg:LSM2}) are designed to handle complex imaging
scenarios with strong velocity contrasts and strong lateral velocity
variations. Examples of such scenarios are salt plays where reflections
underneath the salt are of interest. To demonstrate the viability of our
imaging approach with on-the-fly source estimation in this scenario, we
consider the challenging Sigsbee2A model of size $24.4\times 9.2$ km.
This model contains a large salt body and a number of faults and point
diffractors. To demonstrate the capability of our approach to handel
this challenging situation, we simulate nonlinear data for a marine
acquisition without a free surface. We model $960$ sources in total,
with each shot record being recorded by an array of $320$ receivers with
$25$ m receiver spacing, a maximum offset of $8$ km and a towing depth
of $15$ m. We use a source wavelet with a peak frequency at
$15\text{Hz}$ (see Figure~\ref{fig_source}) and we record for $10$ s.

As is customary during imaging under salt, we use a background velocity
model that features salt with relatively strong and therefore reflecting
boundaries. We approximate linear data by using this background velocity
model to generate data, which we subtract from the simulated data in the
true Sigsbee2A model (i.e.~from the observed data). Due to the presence
of salt in the background model, the incident wavefield contains
reflections that give rise to unwanted tomographic low-frequency
artifacts in the image. This problem is widely reported in the
literature (e.g. \citet{Yoon2006}; \citet{Guitton2007}). To remove these
imaging artifacts, we replace the conventional imaging condition for RTM
by the inverse-scattering imaging condition
\citep{stolk2012linearized, whitmore2012applications, witte2017EAGEspl}.
While this condition has been proven capable of removing tomographic
artifacts during RTM \citep{whitmore2012applications, witte2017EAGEspl}
and sparsity-promoting least-squares RTM \citep{witte2017EAGEspl}, it
changes the linearized forward operator (the Jacobian $\mathbf{J}_i$),
resulting in an inconsistent system. Contrary to RTM with the
conventional imaging condition, imaging with the inverse scattering
imaging condition corresponds to estimating perturbations in the
impedance, rather than in the velocity.

Unfortunately, the difference in which quantity is being imaged, is
problematic for our proposed on-the-fly source estimation, which tries
to correct for inconsistencies between observed and predicted data.
Contrary to the situations where we use the conventional imaging
condition, the data residual now contains contributions from the wrong
wavelet and the linearized imaging condition, which leads to wrong
estimates for the unknown source function. We overcome this problem via
a hybrid iterative algorithm where we switch imaging conditions during
the iterations as outlined in Algorithm~\ref{alg:LSM2}. To estimate the
source function, we first iterate with the conventional imaging
condition. Since the convergence to the source function is fast, we
switch after five iterations to the scattering imaging, but keep the
estimated source function fixed. Basically, we jump from
Algorithm~\ref{alg:LSM2} to Algorithm~\ref{alg:LSM1}.

\begin{figure}
\centering
\subfloat[\label{fig_LSRTM_Sigsbee_a}]{\includegraphics[width=1.000\hsize]{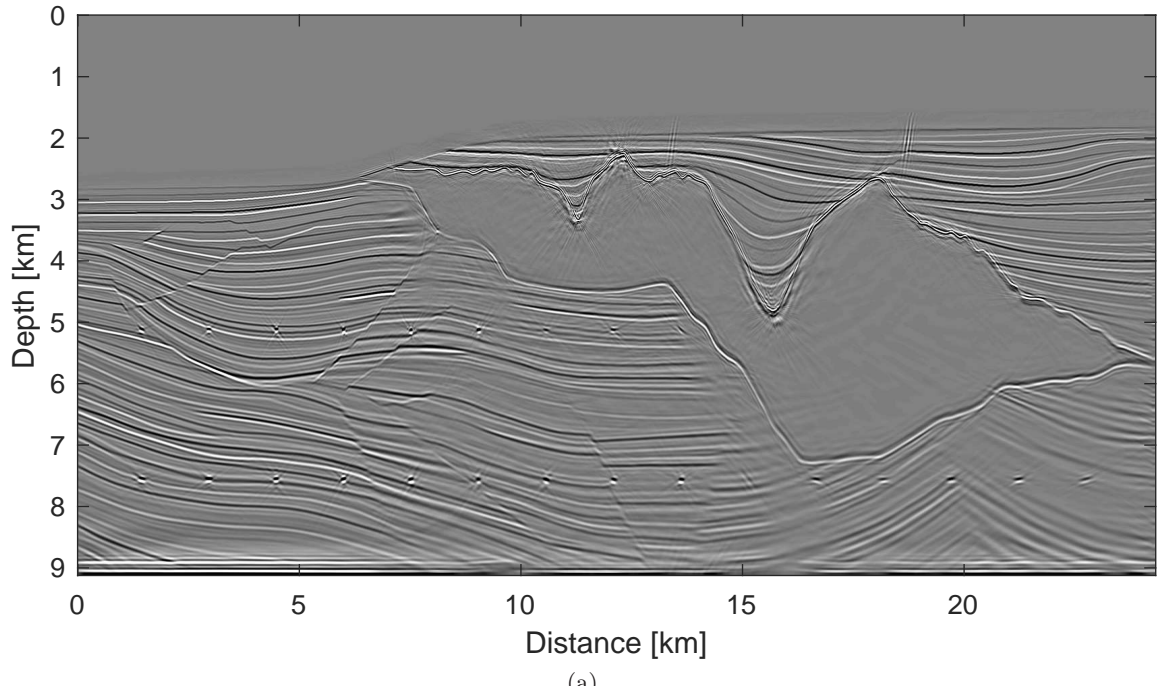}}
\\
\subfloat[\label{fig_LSRTM_Sigsbee_b}]{\includegraphics[width=1.000\hsize]{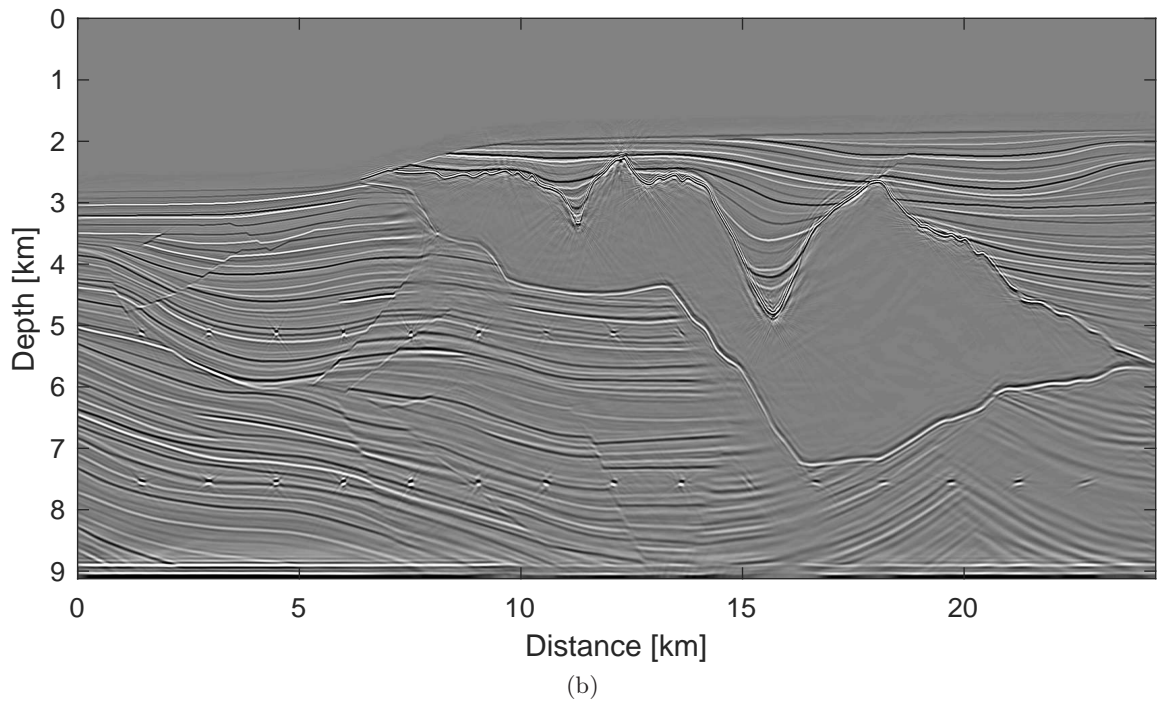}}
\caption{Comparison between (a) SPLS-RTM with the true source wavelet
and (b) SPLS-RTM with on-the-fly source estimation. In both cases, we
conduct total $20$ iterations amounting to two data
passes.}\label{fig_Sigsbee}
\end{figure}

\begin{figure}
\centering
\subfloat[\label{fig_dm_traces_a}]{\includegraphics[width=0.500\hsize]{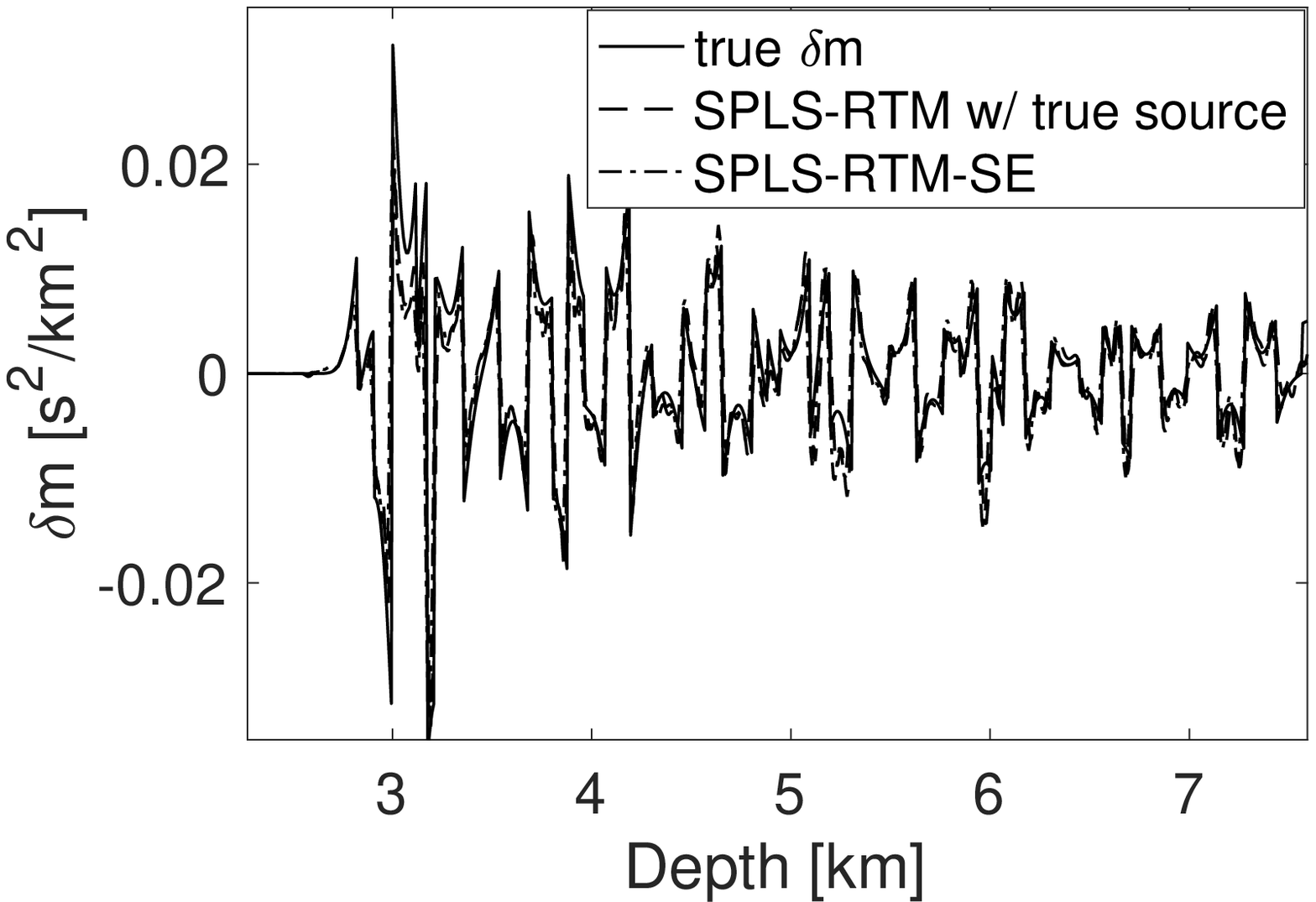}}\hspace*{.4mm}
\subfloat[\label{fig_dm_traces_b}]{\includegraphics[width=0.500\hsize]{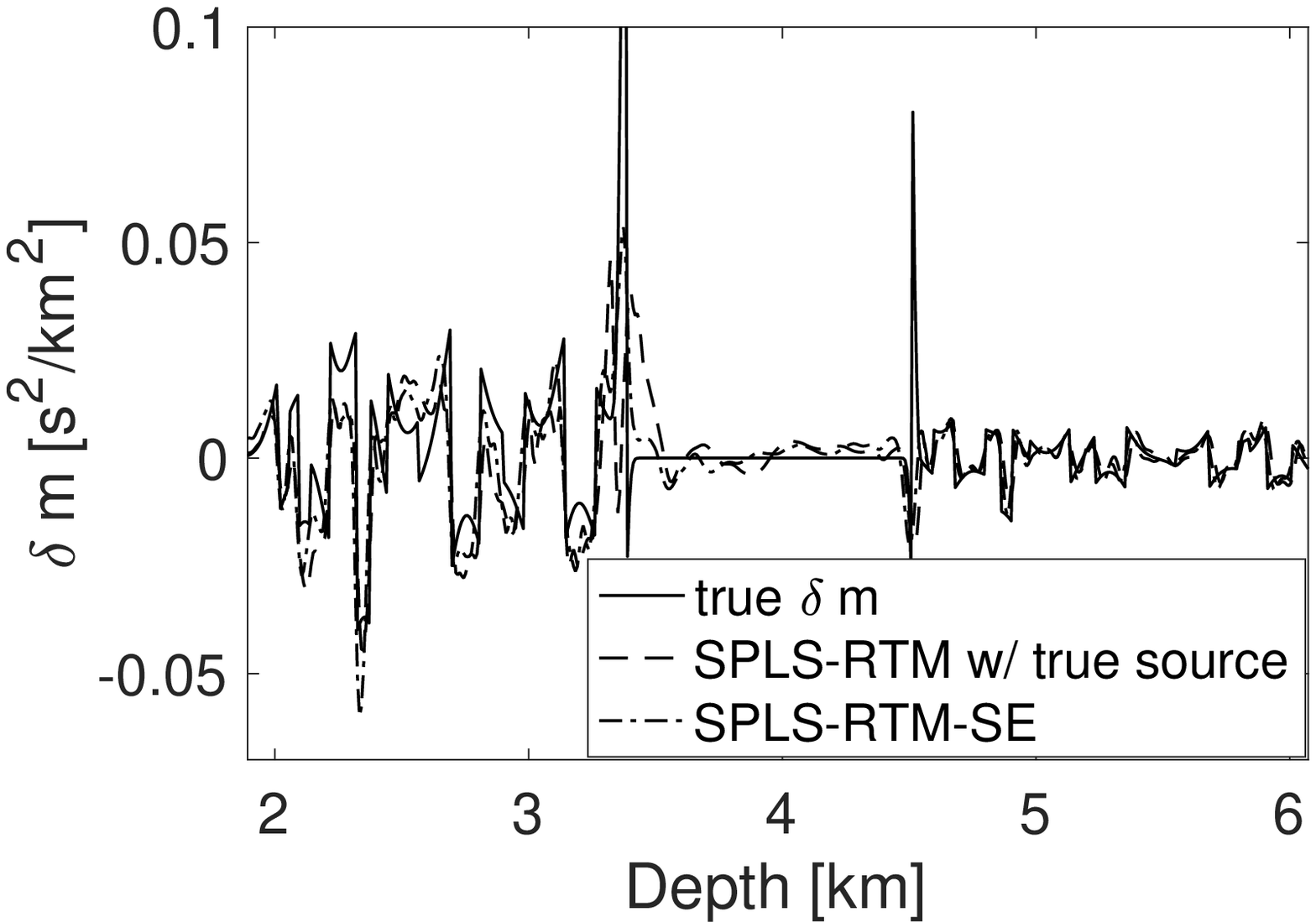}}
\caption{Trace by trace comparisons between the true model perturbation,
the images from SPLS-RTM with the true source and with on-the-fly source
estimation. The traces in (a) and (b) are extracted from lateral
positions $x=4.5$ and $x=11.3$ km, respectively.}\label{fig_dm_traces}
\end{figure}

Results of this hybrid approach are summarized in
Figures~\ref{fig_source} --~\ref{fig_dm_traces}. As before, we compare
our results with on-the-fly source estimation to SPLS-RTM for the true
source function. The initial guess and estimated wavelets in
Figure~\ref{fig_source} again confirm the validity of our approach,
yielding a reasonably accurate estimate for the source after only five
iterations and subsequent normalization of the $\ell_2$-norm. Imaging
results obtained after twenty iterations with $10\%$ of the shots, which
amounts to two data passes in total, are included in
Figures~\ref{fig_Sigsbee} and~\ref{fig_dm_traces}. Unlike a typical RTM
image, images obtained by SPLS-RTM are well resolved and contain true
amplitude. This is because we invert the linearized modeling operator,
which compensates for the source, finite aperture, and propagation
effects. As before, we include preconditioners and mutes to improve the
conditioning number of the linear system. Comparisons of
Figure~\ref{fig_LSRTM_Sigsbee_a}, obtained with Algoritm~\ref{alg:LSM1}
with the true source function, and Figure~\ref{fig_LSRTM_Sigsbee_b},
which we compute with our hybrid method switching from
Algoritm~\ref{alg:LSM2} to Algoritm~\ref{alg:LSM1} after five
iterations, show near identical results, thus confirming the validity of
the proposed approach. These observations are confirmed by
trace-by-trace comparisons in Figure~\ref{fig_dm_traces}. Similar to the
Marmousi experiments, we set the penalty parameters $\alpha=8$, and
$\nu=1$, and the thresholding parameter $\lambda$ is set according to
$10\%$ of the maximum absolute amplitude level of $\mathbf{z}_1$.

\section{Discussion}\label{discussion}

Due to its computational costs, sparsity-promoting LS-RTM has been an
expensive proposition and is for this reason not yet widely adapted
while this method is capable of achieving images with high-amplitude
fidelity and fewer artifacts. With the proposed work, we are able to
come up with an alternative low-cost approach combining techniques from
stochastic optimization and sparsity-promotion with on-the-fly source
estimation using the technique of variable projection. Compared to
earlier work, addressing the memory demands of (LS-)RTM via
on-fly-Fourier transforms \citep{witte2019compressive}, we tackle the
problem of on-the-fly source estimation in the time domain. Because we
use industry-strength time-domain finite-difference propagators provided
by Devito \citep{devito-compiler, devito-api} and exposed in the Julia
programming language by JUDI \citep{witte2018alf}, our approach scales
in principle to large 3D industrial problems. While we address the
importance of estimating source function, we believe that the
sensitivity of LS-RTM to errors in the background velocity model needs
to be studied as well albeit early work on time-harmonic LS-RTM showing
some robustness with respect to these errors \citep{tu2014fis}.
Combining our approach with the method of on-the-fly Fourier transforms
is also a topic that needs further study.

While carrying out full scale 3D (LS-)RTM experiments is generally out
of reach for academia where access to large high-performance compute is
often limited, recent work on a serverless implementation of RTM
\citep{witte2019RHPCssi, witte2019TPDedas} has shown that industry-scale
workloads can be run in the cloud leveraging the power of Devito. In the
not too distant future, we plan to demonstrate the presented method on
an industry-scale imaging problem using tilted transversely-isotropic
propagators. This would truly exemplify the power of modern code bases
and linear algebra abstractions as utilized by JUDI
\citep{witte2018alf}. This framework gives us flexibility for instance
to switch to more involved 3D propagators or to estimating source-time
functions that are allowed to vary along the survey.

Finally, since sparsity-promoting LS-RTM carries out inversions, we
expect to be able to obtain images from sparsely sampled data, e.g.~data
collected with sparse ocean bottom nodes and (multi-)source vessel
simultaneous recordings. We plan to report on these aspects in the not
too distant future.

\section{Conclusion}\label{conclusion}

We proposed a scalable time-domain approach to sparsity-promoting
least-squared reverse time migration with on-the-fly source estimation
in principle suitable for industrial 3D imaging problems. The presented
approach leverages recently developed techniques from convex
optimization and variable projection that greatly reduce costs and the
necessity to provide an estimate for the source function. As a result,
our approach is capable of generating high-fidelity true-amplitude
images including source estimates at the cost of roughly one to two
migrations involving all data.

By means of carefully designed experiments in 2D, we were able to
demonstrate that our method is capable of handling noisy data and
complex imaging settings such as salt. We were able to image under salt,
which is often plagued by low-frequency tomographic artifacts, by
switching between applying the conventional imaging condition initially,
followed by iterations that apply the inverse-scattering condition. In
this way, we estimated the source function first while creating an
artifact-free image with later iterations during which the imaging
condition was switched while keeping the source function fixed.

Because the presented method relies on time-domain propagators, we
anticipate it will be able to scale to large 3D industrial imaging
problems. Because 3D imaging with full-azimuthal sparse data typically
provided good illumination of the reservoir, we expect the proposed
methodology to produce high fidelity results at a cost of roughly one to
two reverse time migrations involving all shots.

\section{Acknowledgements}\label{acknowledgements}

This research was funded by the Georgia Research Alliance and the
Georgia Institute of Technology. And this work is a collaborative effort
of all the co-authors. We confirm that there is no conflict of interest
to declare.

\section{Data availability statement}\label{data_availability_statement}

The data that support the findings of this study are available from the corresponding author upon reasonable request.

\bibliography{SE}

\end{document}